# Multilayer Capacitances: How Selective Contacts Affect Capacitance Measurements of Perovskite Solar Cells

Sandheep Ravishankar,[1*] Zhifa Liu[1], Uwe Rau[1] and Thomas Kirchartz[1,2]

[1]IEK-5 Photovoltaik, Forschungszentrum Jülich, 52425 Jülich, Germany
[2]Faculty of Engineering and CENIDE, University of Duisburg-Essen, Carl-Benz-Str. 199, 47057 Duisburg, Germany

*author for correspondence, email: s.ravi.shankar@fz-juelich.de

**Abstract**

Capacitance measurements as a function of voltage, frequency and temperature are a useful tool to gain deep insight into the electronic properties of semiconductor devices in general and of solar cells in particular. Techniques such as capacitance-voltage, Mott-Schottky analysis or thermal admittance spectroscopy measurements are frequently employed to perovskite solar cells in order to obtain relevant parameters of the perovskite absorber. However, state-of-the-art perovskite solar cells use thin electron and hole transport layers to improve the contact selectivity. However, these contacts are often quite resistive in nature, which implies that their resistance will significantly contribute to the total device impedance and thereby also affect the overall capacitance of the device thereby partly obscuring the capacitance signal from the perovskite absorber. Based on this premise, we develop a simple multilayer model that considers the perovskite solar cell as a series connection of the geometric capacitance of each layer in parallel with their voltage-dependent resistances. Analysis of this model yields fundamental limits to the resolution of spatial doping profiles and minimum values of doping/trap densities, built-in voltages and activation energies. We observe that most of the experimental capacitance-voltage-frequency-temperature data, calculated doping/trap densities and activation energies reported in literature are within the derived cut-off values, indicating that the capacitance response of the perovskite solar cell is indeed strongly affected by the capacitance of its selective contacts.

**Popular summary**

Non-radiative trap-mediated recombination routes are important factors limiting the performance of state-of-the-art perovskite solar cells (PSCs). Identifying the distribution of these trap densities in the perovskite layer is therefore critical to guide device design and passivation strategies. Capacitance techniques are, in principle, effective tools for this purpose. However, we demonstrate that such measurements in the case of PSCs are unreliable because the often-employed thin and low-mobility selective contact layers contribute significantly to the measured capacitance. We show that reported literature data are in several cases artefacts generated by the response of the selective contacts coupled with charge injection, invalidating the current assumption in the perovskite community that suggests the perovskite/selective contact interfaces contain several orders of trap densities higher than the perovskite bulk. We establish resolution limits to avoid these artefacts, which can be applied to any photovoltaic technology that incorporates additional semiconducting layers in addition to the absorber layer.

## 1. Introduction

Perovskite solar cells (PSCs) have developed tremendously over the past decade, with the current highest reported efficiency for a single junction cell standing at 25.5%,[1] close to that of a single junction crystalline silicon solar cell (26.7%).[2] This growth has been possible largely due to structural modifications of the perovskite absorber for high-quality, defect-free films that allow efficient transport of charges, combined with passivation of non-radiative recombination centres in its bulk.[3-6] Consequently, the remaining losses in state-of-the-art PSCs originate mainly from trap-mediated recombination at or close to the interfaces between the perovskite and selective contacts and also resistive losses within the selective contacts.[7-10] Characterization of these loss mechanisms is therefore of vital importance to develop suitable design strategies to further improve the performance of state-of-the-art PSCs.

Capacitance-based techniques have frequently been used for this purpose in perovskite and other emerging solar cell technologies, to measure doping and defect densities as well as recombination coefficients or even mobilities.[11] The general idea is based on the fact that the capacitance is sensitive to the charge stored in the device. If the capacitance is measured as a function of variables such as DC bias voltage, frequency of the AC excitation or temperature, conclusions about various important parameters can be made. For example, identification of the chemical capacitance from capacitance-voltage (CV) measurements can provide information on the much sought-after recombination lifetime, while the identification of the space-charge capacitance related to the existence of a depletion region in the device allows calculating and spatially resolving the doping density in the absorber.[12] Capacitance measurements over several orders of frequency allow determination of the energetic depth and corresponding density of trap states, based on the characteristic (de-trapping) frequency of the trap.[13] This can be combined with temperature and voltage-dependent measurements to determine the distribution of trap states both spatially and energetically in the solar cell in an apparently straightforward manner.[13,14]

However, the validity and applicability of these data interpretation approaches depend on several conditions, one of which is the assumption that the capacitance of the absorber layer (in our case the perovskite layer) either dominates the total capacitance or can be separated from other contributions to the total measured capacitance of the complete device. This is a critical and often invalid assumption, because there are several different capacitances that can respond in a given measurement. The characteristic frequencies or voltage range of response of these different capacitances overlap to a certain degree, making it either extremely difficult or impossible to isolate a specific capacitive contribution that is needed for a given analysis method. Additionally, a capacitance measurement requires the measurement of a current density going into or out of the solar cell, which requires the use of contact layers. Due to the fact that metal/semiconductor interfaces are regions of high surface recombination velocities, selective contacts or transport layers, which selectively allow for the transport of one carrier while blocking the other, are placed between the metal contacts and the absorber layer. In the case of PSCs, these transport layers are either made of metal oxides (e.g. $TiO_2$, $SnO_2$ or NiO) or of organic materials (e.g. PTAA, PCBM, Spiro-OMetad). Especially the organic transport layers have fairly poor mobilities ($10^{-5}$-$10^{-2}$ $cm^2V^{-1}s^{-1}$)[15,16] and are therefore fairly resistive, thin layers of tens of nm thickness that contribute substantially to the impedance, and therefore the capacitance, of the device.

In order to unify the interpretation of the effect of the selective contacts on different capacitance methods, we develop a simple electrical model that considers the perovskite solar cell as a series connection of the geometric capacitances of the perovskite and selective contact layers, each in parallel to its voltage-dependent resistance. This extremely simple model lends itself to an analytical and comprehensive treatment that reproduces a range of non-trivial features that are frequently seen in experimental data. We complement these analytical calculations with more sophisticated numerical frequency-domain drift-diffusion simulations and compare both theoretical approaches with experimental observations from ourselves and other groups using several different capacitance methods such as Mott-Schottky, doping profile and thermal admittance spectroscopy (TAS) measurements. We will show that in all these cases, our equivalent circuit model allows us to define a validity region of the method that depends on material and device parameters such as permittivities and thicknesses. In the case of Mott-Schottky or doping profile measurements at forward bias, this leads to a fundamental limit below which the measured charge densities cannot be considered to originate from dopant or trapped charges. For charge densities below this limit, we show that the built-in voltage obtained from the Mott-Schottky plots is an artefact originating from the voltage-dependent recombination resistance of the perovskite layer, leading to a characteristic dependence on open-circuit voltage and measurement frequency. In the case of TAS measurements, the capacitance of the charge transport layers leads to a minimum activation energy observed that depends on the built-in electrostatic potential drop over the selective contact layers. In addition, we identify that several of these reported parameters in literature are within the derived limits, indicating an urgent need to revise the information obtained regarding PSCs from capacitance measurements.

## 3. Results and discussion

The key challenge in analysing solar cell capacitance measurements is unravelling the superposition of different effects that contribute to the capacitance. Any attempt to analyse the data requires knowing and considering the different contributions to the capacitance. We therefore begin by briefly highlighting the main features of some fundamental capacitances that are observed experimentally and theoretically in solar cells (extended discussion provided in section A1 of the supporting information). These comprise of three capacitances – the depletion capacitance, the chemical capacitance and the diffusion capacitance. We also clarify that all capacitances referred to in this paper indicate a capacitance per unit area (Fcm$^{-2}$).

The depletion capacitance is associated to a space-charge region of width $w$ in the absorber layer of thickness $d$, where the majority-carrier concentration is negligible with respect to the bulk. This occurs as a consequence of the electrostatic potential drop across this region due to a density of dopants or traps. This region acts as a parallel-plate capacitor whose plate spacing is voltage-dependent, leading to its capacitance having an inverse square root dependence ($C_{sc} \propto 1/\sqrt{V_{bi} - V}$) on the applied voltage $V$ measured with respect to the built-in voltage $V_{bi}$. This means that a plot of $C_{sc}^{-2}$ versus $V$ approaches a straight line in the voltage region where the space-charge-region capacitance dominates. The plot of $C_{sc}^{-2}$ versus $V$ is termed a Mott-Schottky plot.[17] The slope of the Mott-Schottky region contains important information regarding the layer such as its doping density and relative permittivity. Its intercept on the voltage axis yields the built-in electrostatic potential difference $V_{bi}$ (see equation S3 in the supporting information). When the absorber layer is fully depleted, the depletion capacitance saturates to the voltage-independent geometric capacitance $C_g$ of the layer.

The chemical capacitance $C_\mu$ and closely related diffusion capacitance (see sections A1.2 and A1.3 in the supporting information), on the other hand, are related to the injection of free charge carriers into the semiconductor layer, which leads to an exponential increase in the capacitance with applied voltage ($C \propto \exp(qV/mk_BT)$), where $m$ is a dimensionless factor that we discuss in section 3.2. In addition to the chemical capacitance of free carriers, there also exists the chemical capacitance of trapped carriers, related to filling of the trap density of states. This capacitance makes a peak at half-occupation of the trap energy level (equation S7 in the supporting information).[18] Another contribution is from the geometric or electrode capacitances of the different layers that comprise the solar cell. In general, the response from these different capacitances can overlap as a consequence of their similar magnitudes of the capacitances, or similar time constants that depend also on their associated resistances. Therefore, we will develop a simple electrical model of the perovskite solar cell that accounts for only the fundamental mechanisms of charge injection and geometric capacitances of each individual layer of the solar cell, with the aim of identifying to which extent these factors affect or overlap with the expected response of the fundamental capacitances.

*3.1. Multilayer capacitances*

The derivation of the fundamental capacitances described previously assumes that the contacts for charge injection/extraction are metals or thin semiconductor layers with sufficient conductivity to have very low transport resistance i.e. to be metal-like. This means that the effective capacitance of the contacts is infinitely large and hence does not make any contribution to the measured capacitance of the absorber layer. However, this assumption is not valid in the case of perovskite solar cells, which frequently use thin organic layers as selective contacts whose mobilities are low compared to those of the perovskite absorber. In such cases, the PSC cannot be considered as a capacitive perovskite layer sandwiched between two metal-like contacts, but rather as three individual layers with their own capacitances.

This approach is illustrated in figure 1(a) and 1(b), where the PSC is modelled as a series connection of the geometric capacitances of each layer. Each of the geometric capacitances $C_{g,\text{layer}}$ is placed in parallel with its corresponding voltage-dependent resistance $R_{\text{layer}}$. We note that the use of passive electrical elements maintains the generality of the model, allowing it to be used for both n-i-p and p-i-n type devices. However, as a focussed point of reference, we use the parameters (see table S2 in the supporting information) of a p-i-n type ITO/PTAA/CH$_3$NH$_3$PbI$_3$/PCBM/Ag device as shown in figure 1(a). The net impedance of this system is

$$Z(V,\omega) = R_s + \left(\frac{1}{R_{\text{PCBM}}(V)} + i\omega C_{g,\text{PCBM}}\right)^{-1} + \left(\frac{1}{R_{\text{pero}}(V)} + i\omega C_{g,\text{pero}}\right)^{-1} + \left(\frac{1}{R_{\text{PTAA}}(V)} + i\omega C_{g,\text{PTAA}}\right)^{-1}. \quad (1)$$

The net capacitance is given by

$$C(V,\omega) = \frac{\text{Im}(Z^{-1})}{\omega} = \text{Im}\left[\frac{1}{\omega}\left(R_s + \frac{R_{\text{PCBM}}(V)}{1+i\omega R_{\text{PCBM}}(V)C_{g,\text{PCBM}}} + \frac{R_{\text{pero}}(V)}{1+i\omega R_{\text{pero}}(V)C_{g,\text{pero}}} + \frac{R_{\text{PTAA}}(V)}{1+i\omega R_{\text{PTAA}}(V)C_{g,\text{PTAA}}}\right)^{-1}\right]. \quad (2)$$

From equations 1 and 2, we note that the impedance and consequently the capacitance is a function of both voltage and frequency. While the frequency dependence of the net capacitance is obvious, its voltage dependence is a consequence of the voltage-dependent resistances of one or more of the layers.

In order to complete the model, we require knowledge of the resistances of the individual layers as a function of voltage. In the case of the perovskite layer, we assume that it is a recombination resistance $R_{\text{rec}}$, which can be easily obtained by differentiating the diode equation as

$$R_{\text{pero}} = \left(\frac{dj_{\text{rec}}}{dV}\right)^{-1} = \frac{n_{\text{id}}k_BT}{qj_0\exp(qV/n_{\text{id}}k_BT)}, \quad (3)$$

where $j_{\text{rec}}$ is the recombination current density, $n_{\text{id}}$ is the ideality factor and $j_0$ is the reverse saturation current. $j_0$ can be calculated using the short-circuit current density $j_{\text{sc}}$ and the open-circuit voltage as

$$j_0 = j_{\text{sc}}\exp\left(\frac{-qV_{\text{oc}}}{k_BT}\right). \quad (4)$$

In order to determine the resistances of the transport layers, we first calculate the dark current through a transport layer. Figure S5 in the supporting information shows the energetics of a perovskite solar cell with the absorber in contact with its respective transport layers and metal electrodes. We assume a constant Fermi-level splitting $V_{\text{int}}$ in the intrinsic perovskite layer, and the difference of Fermi levels of electrons and holes at the transport layer/electrode interfaces is determined by the external applied voltage $V_{\text{ext}}$. We also assume a constant built-in electrostatic potential $V_{\text{bi,TL}}$ through each transport layer. The dark current through the transport layer is then given by (see section A2 in the supporting information for derivation)

$$j = j_{0,\text{TL}}\left(\exp\left(\frac{qV_{\text{TL}}}{k_B T}\right) - 1\right), \tag{5}$$

where $j_{0,\text{TL}}$ is the dark current pre-factor and $V_{\text{TL}}$ is the potential drop across the transport layer. The transport layer resistance $R_{\text{TL}}$ can be calculated analytically (see section A2 in the supporting information) and is given by

$$R_{\text{TL}} \cong \frac{d}{q\mu n_0 \left[\frac{q(V_{\text{bi,TL}} - V_{\text{el,TL}})}{k_B T}\right]} \left[\exp\left(\frac{q(V_{\text{bi,TL}} - V_{\text{el,TL}})}{k_B T}\right) - 1\right] + \frac{dk}{2q\mu n_0}, \tag{6}$$

where $V_{\text{el,TL}}$ is the amount of the external voltage that goes to alter the electrostatic potential (controlled by a parameter $k$) of the transport layer, $d$ is the thickness of the transport layer, $\mu$ is the mobility of the majority carriers and $n_0$ is the majority carrier concentration at the transport layer/electrode interface determined by the injection barrier.

Plugging the resistance in equations 6 into equation 2, we can calculate the evolution of capacitance for this model versus both voltage and frequency. We first discuss its evolution versus frequency, shown in figure 1(c), to provide a basic understanding of the model.

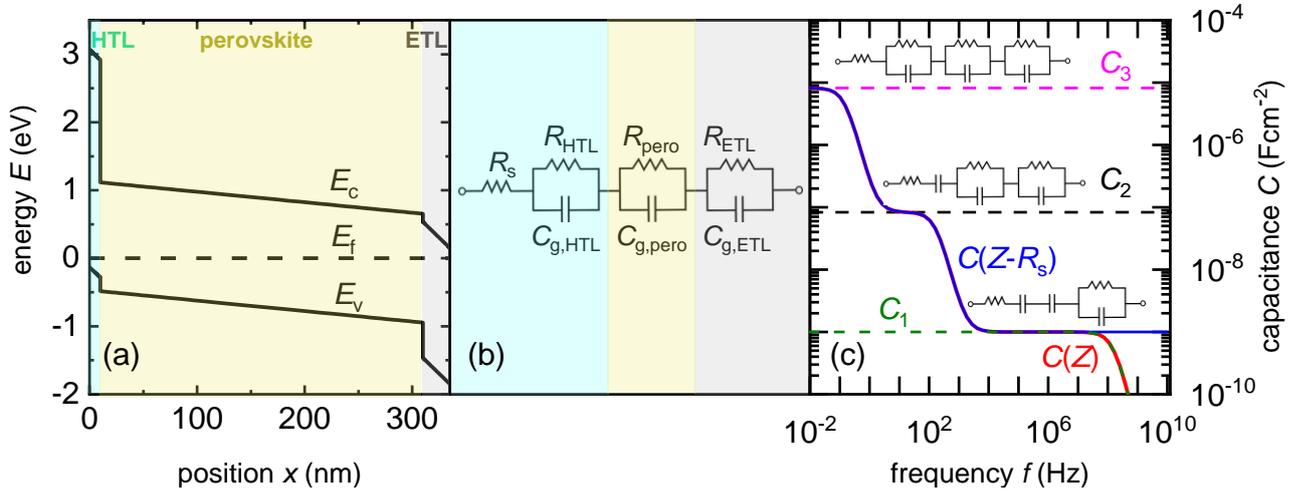

**Figure 1** (a) Band diagram of the reference perovskite solar cell upon which the simulations of this paper are based (see table S1 for parameters). (b) Equivalent circuit model of the solar cell in the dark. $R_s$, $R_{\text{HTL}}$, $R_{\text{pero}}$ and $R_{\text{ETL}}$ are the series resistance and voltage-dependent resistances of the individual layers respectively, and $C_{g,\text{HTL}}$, $C_{g,\text{pero}}$ and $C_{g,\text{ETL}}$ are the geometric capacitances of the individual layers respectively. (c) Simulated capacitance evolution versus measurement frequency for the equivalent circuit of (b), with ($C(Z)$) and without ($C(Z - R_s)$) a series resistance, for well separated time constants of each $R||C$ element. Three plateaus in the capacitance are observed at high frequency ($C_1$), intermediate frequency ($C_2$) and low frequency ($C_3$), whose analytical solutions are derived in section A3 in the supporting information and shown in (c). The corresponding equivalent circuit representing the capacitance plateau is shown in the inset next to each plateau. Note that the high-frequency drop in capacitance is simply a consequence of the series resistance as derived in section A3 in the supporting information.

In case of well-separated time constants of each $R||C$ pair, we observe three capacitance plateaus, one at high frequency ($C_1$), one at intermediate frequency ($C_2$) and one at low frequency ($C_3$). We hereafter refer to the three capacitances and their corresponding resistances using the subscripts 'HF', 'IF' and 'LF' corresponding to the magnitude of their characteristic frequencies (high, intermediate and low frequency) and the subscripts 1-3 for the capacitance plateaus observed. The analytical equations (see section A3 in the supporting information for the derivation) of the capacitance plateaus are

$$C_1 = \left[\frac{1}{C_{\text{HF}}}\left(\frac{R_\text{s}+R_{\text{HF}}}{R_{\text{HF}}}\right)^2 + \frac{1}{C_{\text{IF}}} + \frac{1}{C_{\text{LF}}}\right]^{-1} \cong \frac{1}{C_{\text{HF}}} + \frac{1}{C_{\text{IF}}} + \frac{1}{C_{\text{LF}}}, \tag{7}$$

$$C_2 = \left[\frac{1}{C_{\text{LF}}} + \frac{(R_\text{s}+R_{\text{HF}}+R_{\text{IF}})^2}{R_{\text{HF}}^2 C_{\text{HF}} + R_{\text{IF}}^2 C_{\text{IF}}}\right]^{-1}, \tag{8}$$

and

$$C_3 = \left[\frac{(R_\text{s}+R_{\text{HF}}+R_{\text{IF}}+R_{\text{LF}})^2}{R_{\text{HF}}^2 C_{\text{HF}} + R_{\text{IF}}^2 C_{\text{IF}} + R_{\text{LF}}^2 C_{\text{LF}}}\right]^{-1}. \tag{9}$$

The effective equivalent circuit representations of these capacitance plateaus are also shown in figure 1(c).

*3.2. Mott-Schottky plots and doping densities*

One of the most important applications of capacitance measurements is the use of the Mott-Schottky plot (equation S3 in the supporting information) to determine the doping density (dopants or traps) in the semiconductor layer. The doping density is calculated from the Mott-Schottky relation as

$$N_\text{d}(V) = \frac{-2(dC^{-2}/dV)^{-1}}{q\epsilon_\text{r}\epsilon_0}, \tag{10}$$

where $N_\text{d}$ is a constant value if the Mott-Schottky region is a straight line. In the case of a spatially non-uniform doping density, $N_\text{d}$ would slightly change with voltage. Within the framework of the depletion approximation, the $N_\text{d}(V)$ would represent the doping density at the edge of the space charge region present at an applied voltage $V$.[19] Thus, in order to determine the doping profile, the convention is then to plot equation 10 versus the distance of the edge of the space charge region to the metal contact that carriers the counter charge to the charge of the majority carriers in the bulk (e.g. in a p-type semiconductor, it would be the distance to the electron-injecting contact). This distance is typically called the profiling distance $w$. The profiling distance follows from a simple plate-capacitor approximation of the depletion region given by

$$w(V) = \frac{\epsilon_\text{r}\epsilon_0}{C(V)}. \tag{11}$$

The spatial resolution of this method is limited by a few Debye lengths, which reflects the fact that the majority carrier density cannot abruptly transition from zero to the doping density value in the neutral zone at the edge of the depletion layer.[19] In addition, in the case of PSCs, the measurement frequency must be chosen carefully so as to avoid capacitive contributions from the slow-moving ions, whose relevant transport timescales are on the order of 0.01-100 Hz.[20] While ionic influences have been suggested even at higher frequencies under illumination,[21] we focus on dark measurements and assume that frequencies an order or two higher than these values are sufficient to neglect ionic contributions to the spectra.

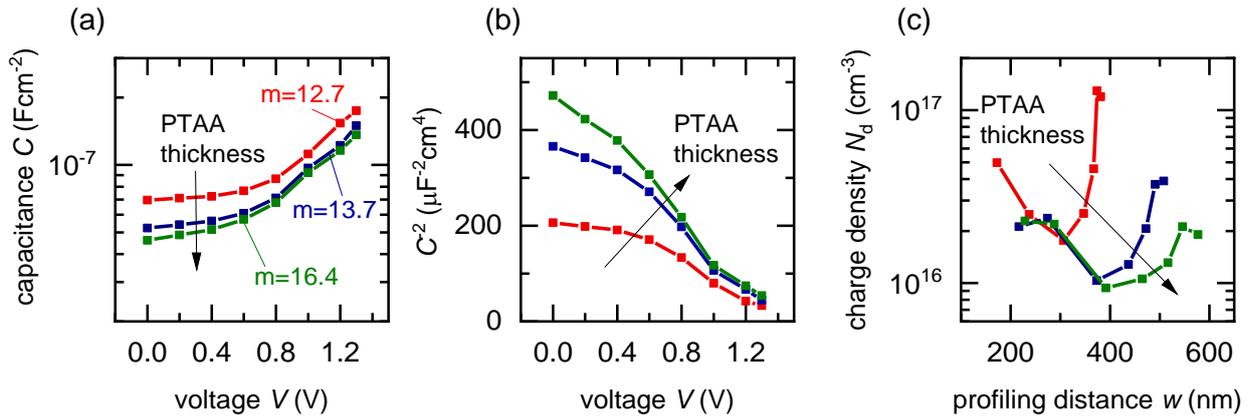

**Figure 2** Experimentally measured (a) capacitance versus voltage plots, (b) Mott-Schottky plots and (c) doping profiles for ITO/PTAA/CH$_3$NH$_3$PbI$_{3-x-y}$I$_x$Br$_y$/PCBM/Ag samples with increasing thickness of the PTAA layer as indicated in the inset. Fits of the capacitance versus voltage to the equation $C = C_g + C_0 \exp(qV/mk_BT)$ to obtain the $m$ value shown in (a) are shown in figure S7 in the supporting information. The measurement frequency was 25469 Hz.

Figure 2(a) shows some experimentally measured capacitance-voltage plots under dark conditions and figures 2(b), (c) show the corresponding Mott-Schottky and doping profile plots respectively for ITO/PTAA/CH$_3$NH$_3$PbI$_{3-x-y}$I$_x$Br$_y$/PCBM/Ag samples with increasing thickness of the PTAA layer. The measurement frequency of 25469 Hz was chosen because it corresponds to the high-frequency plateau in the capacitance, which should be related to the free carrier response. The capacitance-voltage profiles show a weak exponential increase of the capacitance starting at around 0.5 V with slope factor $m$ between 12.7 and 16.4 (see figure S7 in the supporting information for fits), making it difficult to identify its mechanistic origin. The corresponding Mott-Schottky plots also show the onset of the linear Mott-Schottky region at intermediate forward bias, around 0.4 V for all the samples.

A summary of Mott-Schottky plots of different perovskite solar cells reported in literature are shown in figure 3(a). The nature of these plots is very similar to that of figure 4(b), showing the linear Mott-Schottky region only at intermediate to large forward bias with either a plateau or weak evolution at lower voltages. In addition, we can also observe a second linear region with reduced slope (downward kink) at large forward bias in some of the plots, which has generally been assigned to a contribution from the exponential chemical or ionic capacitance.[22,23] The literature data suggests doping densities in the perovskite layer (calculated from their corresponding doping profiles, see figure S7(b) in the supporting information) between $3 \times 10^{15} - 7 \times 10^{17}$ cm$^{-3}$. As a reference, a Mott-Schottky plot of a mono-crystalline silicon solar cell is also shown in figure 3(a), which shows a similar doping density to the perovskite data points but a very different Mott-Schottky plot, yielding a clear linear Mott-Schottky region from deep reverse bias to forward bias. SCAPS simulations of perovskite solar cells with different doping densities shown in figure 3(b) also indicate that the onset of the Mott-Schottky region should occur from deep reverse bias, while conversely, for doping densities ~$10^{15}$ cm$^{-3}$, the behaviour is identical to that of an intrinsic perovskite. Furthermore, the expected onset voltage $V_{\text{onset}}$ of the space-charge capacitance is plotted as a function of doping density and perovskite layer thickness in figure 4(a), calculated by setting the space-charge capacitance equal to the geometric capacitance of the perovskite layer. For typical thin film perovskite solar cells of thickness 300 nm, doping densities in excess of $\cong 2 \times 10^{16}$ cm$^{-3}$ should show an onset of the Mott-Schottky plot at reverse bias, which is not observed in the literature data in figure 3(a). However, for lower reported doping densities, the onset voltage coincides with that observed in the literature data i.e. intermediate to large forward bias.

Therefore, based on the analysis so far, it is still difficult to confirm the validity of the doping densities reported in literature from the Mott-Schottky analysis of perovskite solar cells.

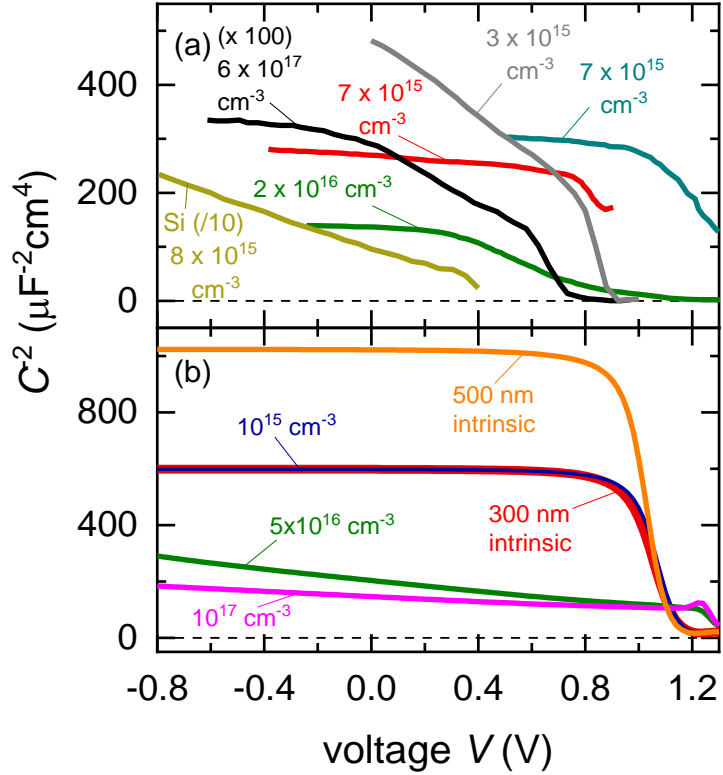

**Figure 3** (a) Mott-Schottky plots of different perovskite solar cells and a silicon solar cell reported in literature. Their corresponding calculated doping densities are shown in the label, obtained from their corresponding doping profiles (see figure S7(b) in the supporting information). (b) Simulated Mott-Schottky plots of a perovskite solar cell with different acceptor doping densities of the perovskite layer as shown in the label, using SCAPS. The orange and red lines are simulations of an intrinsic perovskite layer with different thicknesses. The simulated Mott-Schottky plots for an intrinsic layer and dopant density of $10^{15}$ cm$^{-3}$ are very similar to each other and to the experimental plots shown in (a). The literature data corresponds to refs.[22-27].

However, ref. [28] has previously discussed the sensitivity of the Mott-Schottky analysis for organic solar cells, identifying that in cases where the doping density is not high enough to alter the bulk carrier concentration sufficiently to create a neutral region, an apparent Mott-Schottky plot similar to the one generated by an intrinsic perovskite shown in figure 3(b) is obtained. This apparent Mott-Schottky region can originate from either the chemical capacitance, or from transitions between the geometric capacitances of the different layers comprising the solar cell, based on the model developed in section 3.1 (see figure S8 in the supporting information). This is because the resistances in this model depend exponentially on the external voltage and hence the capacitance also shows an exponential evolution with the external voltage (see equations 2,3 and 6). To account for the effects of these two additional voltage-dependent capacitances that can create an apparent Mott-Schottky region, we define the general voltage-dependent capacitance as

$$C(V) = C_\mathrm{g} + C_0 \exp(\frac{qV}{mk_\mathrm{B}T}). \tag{12}$$

where $C_g$ is the geometric capacitance, $C_0$ is a pre-factor and $m$ is a parameter that controls the slope of the exponential evolution of the capacitance versus voltage. Carrying out a Mott-Schottky analysis of equation 12 using equation 10 (see sections A4 and A5 in the supporting information for the derivation), we obtain a charge density given by[29]

$$N_{\text{d,min}} = \frac{27 m k_B T \epsilon_r \epsilon_0}{4 q^2 d_{\text{pero}}^2} . \qquad (13)$$

Equation 13 provides a lower limit of resolution and suggests that only charge densities that are significantly above this limit can be considered as originating from doping or trap densities in capacitance measurements. For commonly used perovskite layer thicknesses between 300-1000 nm, we obtain apparent doping densities between $6 \times 10^{15}$ cm$^{-3}$ and $2 \times 10^{14}$ cm$^{-3}$, assuming $m = 1$. The large values of $m$ observed experimentally ($m > 10$, see figure 2(a)) then suggest that this limit increases by at least an order of magnitude to $6 \times 10^{16}$ cm$^{-3}$ for a 300 nm thick perovskite layer and $2 \times 10^{15}$ cm$^{-3}$ for a 1000 nm thick perovskite layer. These charge densities are very similar to the measured doping densities seen in figure 3(a), suggesting that the calculated doping densities are erroneous and in fact originate from the response of the chemical capacitance or geometric capacitances, rather than from a depletion layer capacitance. In addition, equation 13 implies that for measured charge densities close to the resolution limit, thin films will always show higher apparent doping or trap densities compared to a bulk single crystal due to the inverse square thickness dependence. Therefore, care must be taken during such measurements since such a trend is intuitively expected based on the higher crystallinity of bulk single crystals compared to thin films.

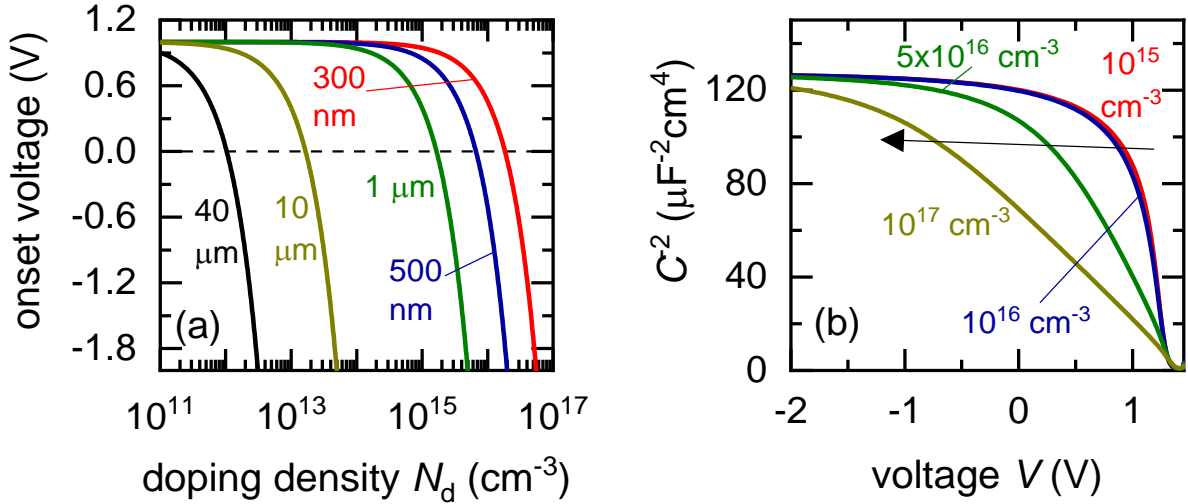

**Figure 4** (a) Onset voltage of the space-charge capacitance as a function of doping density for different thicknesses of the absorber layer, calculated by setting the space-charge capacitance (equation S1 in the supporting information) equal to the geometric capacitance. For thin film solar cells, dopant densities in excess of $10^{16}$ cm$^{-3}$ are required to observe the space-charge capacitance at reverse bias. (b) SCAPS simulations of Mott-Schottky plots of a perovskite solar cell (without selective contacts and 0 eV injection barrier on either side) with different doping densities of the perovskite layer as shown in the label. The arrow indicates the shift in the onset voltage of the linear Mott-Schottky region with increasing doping density.

We now justify the use of equation 12 to represent the capacitance-voltage evolution of the multilayer model developed in section 3.1. Figure 5(a) shows the simulated capacitance evolution of this model versus applied voltage and figure 5(b) shows the corresponding Mott-

Schottky plot. The capacitance remains flat through most of the voltage range before it starts to increase exponentially and peaks at a certain value. This creates a Mott-Schottky plot very similar to the ones observed in figure 3(a). To understand how the constant geometric capacitances of the perovskite solar cell can make a step in the capacitance versus voltage, we must look at the evolution of the characteristic frequencies of the layers and their corresponding resistances. The characteristic frequencies are given by

$$\omega_{layer}(V) = [R_{\text{layer}}(V)C_{\text{layer}}]^{-1}, \tag{14}$$

where the 'layer' subscript indicates the name of the different layers comprising the solar cell. The voltage dependence of the resistances of the PTAA, perovskite and PCBM layers (see equations 3 and 6) causes their corresponding characteristic frequencies to evolve with voltage, both of which are plotted in figure 5(c) and figure 5(d) respectively. The resistance of the perovskite layer decreases exponentially with applied forward bias due to the fact that it is a recombination resistance, while the contact layer resistances show an exponential decrease with applied forward bias related to increasing conductivity upon injection of majority carriers and saturates at a minimum value $R_{\text{TL,min}}$ (see equation S37 in the supporting information). By comparing the values of the characteristic frequencies to the measurement frequency $\omega_{\text{exp}}$, we can derive analytical approximations (see section A3 in the supporting information) of the low and high voltage plateaus of the capacitance ($C_{\text{LV}}$ and $C_{\text{HV}}$ respectively) in figure 5(a) as

$$C_{\text{LV}} = \left(\frac{1}{C_{\text{ptaa}}} + \frac{1}{C_{\text{pero}}}\right)^{-1}, \tag{15}$$

$$C_{\text{HV}} = C_{\text{ptaa}}, \tag{16}$$

with the principle being that only those capacitances whose characteristic frequencies are much smaller than the measurement frequency dominate the net capacitance. The low voltage plateau is then the series connection of the geometric capacitances of the PTAA and perovskite layers, which transitions to a high voltage plateau of the geometric capacitance of the PTAA layer. Comparing the Mott-Schottky behaviour of the multilayer model with that of the depletion capacitance and the chemical capacitance shows that their behaviour is very similar and overlaps at the same voltage range, with only reverse bias conditions being suitable to clearly isolate the response of the depletion capacitance (see figure S8 in the supporting information).

Therefore, in order to discriminate such effects from an actual Mott-Schottky behaviour arising from a depletion layer capacitance, we must ensure the calculated doping densities are significantly higher than equation 13. Additionally, we note that in any working solar cell, there has to be some amount of charge injection and hence a contribution from the chemical capacitance to the total capacitance (which might be hidden, as is the case for perovskite solar cells). This suggests that the capacitance evolution of the perovskite solar cell is likely a combination of the chemical capacitance and multilayer geometric capacitance transition mechanisms, both of which evolve exponentially, as previously discussed. Simulations including the chemical capacitance in the multilayer capacitance model are shown in figure S9 in the supporting information. Larger $m$ values alter the plateau at low voltages, creating a weak evolution into the apparent Mott-Schottky region, similar to that observed experimentally in figure 3(a). Note that the downward kink at large forward bias seen in the Mott-Schottky plots in figure 3(a) is also reproduced in this model.

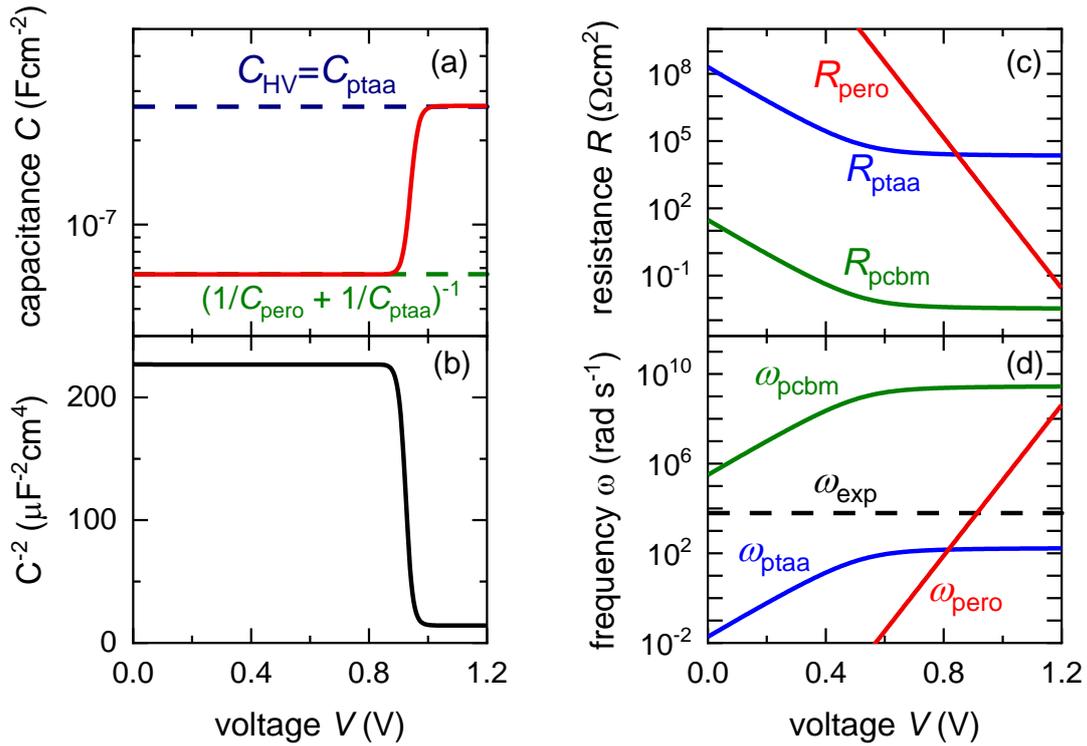

**Figure 5** Simulated evolution of (a) capacitance, (b) corresponding Mott-Schottky plot, (c) resistances and (d) characteristic frequencies versus applied voltage based on the multilayer model developed in section 3.1. Parameters are discussed in table S2 in the supporting information. The green and blue dashed lines in (a) correspond to the low voltage capacitance limit $C_{LV}$ and high voltage capacitance limit $C_{HV}$ (equations S50 and S52 in the supporting information). The measurement frequency was chosen as $f_{exp} = 10^3$ Hz, which is a typical frequency for Mott-Schottky measurements of perovskite solar cells that is assumed fast enough to avoid direct ionic influences to the spectra.

*3.3. The relationship between the open-circuit voltage and built-in voltage*

Another important parameter that is obtained from Mott-Schottky plots (intercept on the voltage axis) is the built-in electrostatic potential drop $V_{bi}$ (generally referred to as the built-in voltage) in the absorber layer. The built-in voltage is generated upon equilibration of the Fermi levels of the metallic and semiconductor layers in the device and depending on the doping density, can drop in a small layer at an interface, as is the case for strongly doped crystalline silicon solar cells, or drop across the entire thickness of the absorber for an intrinsic solar cell. In general, the built-in voltage is important to maximise the performance of a solar cell, offering the beneficial properties of charge selectivity and preventing the formation of extraction barriers for electrons and holes at large forward biases. For intrinsic, low-mobility semiconductors, the $V_{bi}$ also allows efficient charge separation and transport at short-circuit conditions while also reducing recombination at open-circuit conditions by reducing the density of minority carriers at either interface.[30,31]

Therefore, it is intuitively expected that larger $V_{bi}$ values will result in higher open-circuit voltages. This is indeed observed from experimental Mott-Schottky data reported in literature, shown in figure 6(a). These data points correspond to cases where the improved performance and open-circuit voltage of a passivated sample compared to an unpassivated sample is interpreted as occurring due to efficient charge extraction as a consequence of the increased $V_{bi}$ observed for the passivated sample from the Mott-Schottky plot. However, the exact mechanism regarding how passivation alters the equilibrium electrostatic potential distribution is unclear.

To confirm the validity of this interpretation, we simulated the effect of variations in open-circuit voltage using the multilayer model, due to the fact that the recombination resistance in the model creates an open-circuit voltage dependence (equations 3 and 4). Figure 6(b) shows the simulated Mott-Schottky plots, which show the same trend of increased apparent $V_{bi}$ with increasing open-circuit voltage as observed experimentally in figure 6(a). An analytical equation showing the open-circuit voltage and frequency dependence of the capacitance step that yields the apparent $V_{bi}$ from the multilayer model is provided in section A7 in the supporting information (equation S79).

We can conclude that in situations where the Mott-Schottky transition does not arise from a depletion capacitance, the observed $V_{bi}$ is only an artefact generated by a step in the capacitance. Therefore, any increase in the open-circuit voltage due to passivation or material modification causes an increase in the apparent $V_{bi}$ and not the other way around. This occurs due to the fundamental dependence of the recombination resistance of the perovskite layer on its open-circuit voltage in combination with charge injection or multilayer capacitance transitions. Arguments of increased charge extraction due to an apparent increased $V_{bi}$ from Mott-Schottky plots therefore cannot be invoked to justify an increase in the open-circuit voltage of passivated samples compared to unpassivated ones.

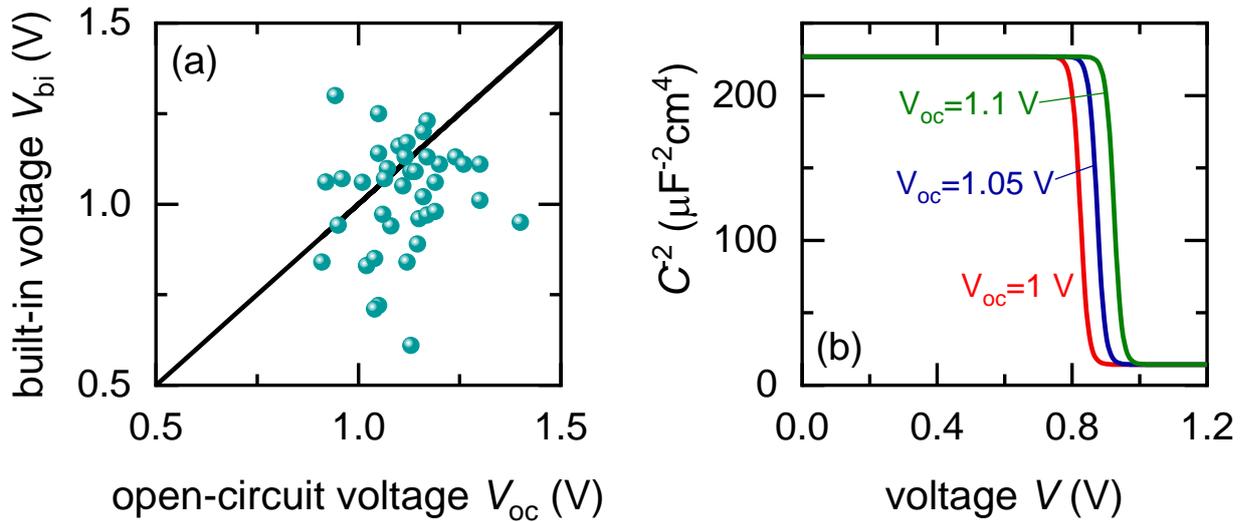

**Figure 6** (a) Plot of reported built-in voltages obtained from Mott-Schottky analysis in literature versus corresponding open-circuit voltages. The solid line indicates the region where $V_{bi} = V_{oc}$. Larger built-in voltages are associated to larger open-circuit voltages. (b) Simulated Mott-Schottky plots for different open-circuit voltages shown in the label, using the multilayer model. The measurement frequency was chosen as $f_{exp} = 10^3$ Hz. The multilayer model predicts the trend observed in (a) for the evolution of built-in voltage versus open-circuit voltage. The literature data corresponds to references [22-25,32-46]. The variation of the built-in voltage of the multilayer model as a function of frequency is shown in figure S12(b) in the supporting information.

*3.4. Interpretation of doping profiles*

We now turn to understanding the spatial dependence of the apparent doping density from doping profile plots. Figure 2(c) shows the measured doping profiles of ITO/PTAA/CH$_3$NH$_3$PbI$_{3-x-y}$I$_x$Br$_y$/PCBM/Ag solar cells for different thicknesses of the PTAA layer. The doping profile makes a 'U' shape, with a plateau region in the apparent bulk of the perovskite layer and steep rises at both small and large apparent profiling distances corresponding to regions close to the interfaces of the perovskite with the selective contacts. This 'U' shape moves to lower charge densities and larger profiling distances for increasing thickness of the PTAA layer. Such a doping profile and its evolution with PTAA layer thickness is intuitively puzzling, however, based on our previous analysis of the Mott-Schottky plots, we speculate that this profile is generated by the combination of the multilayer model and the chemical capacitance. To confirm our hypothesis, we carried out SCAPS simulations of spatial doping profiles of thin film, trap-free, dopant-free perovskite solar cells with different thicknesses of the perovskite layer, shown in figure 7(a).

All the profiles form the 'U' shape observed in figure 2(c) even in the absence of any traps or dopant densities. The simulated apparent charge densities of a perovskite solar cell with 300 nm thickness of the perovskite layer is similar in magnitude to the experimental values seen in figure 2(c). Furthermore, an analysis of the doping profiles of several different photovoltaic technologies shows that they all yield the same 'U' shape,[29] which would be quite a coincidental occurrence. These observations suggest that these apparent charge densities are indeed related to signals from the chemical capacitance and the multilayer model. We therefore derive analytical approximations (see sections A4 and A5 in the supporting information) of the doping profile by considering a simple geometric capacitance in parallel to a general capacitance that evolves exponentially with voltage, as carried out previously (equation 13). Since the doping profile is determined by the inverse slope of the Mott-Schottky plot (equation 10), the rise in apparent charge density at large profiling distances (reverse bias) is simply due to the constant geometric capacitance. The plateau region in the doping profile thus corresponds to the apparent linear Mott-Schottky region in the bulk of the sample (intermediate forward bias), yielding a minimum doping density value. For small profiling distances (large forward bias), we observe another sharp rise in doping density corresponding to the plateau region at large forward bias of the Mott-Schottky plot. The plateau region and the forward bias rise of the doping profile can be analytically calculated as[29]

$$N_\text{d}(w) = N_{\text{d,min}} + \frac{mk_\text{B}T\epsilon_r\epsilon_0}{q^2w^2}, \qquad (17)$$

where $N_{\text{d,min}}$ is the minimum charge density shown in equation 13. A simulated doping profile using the multilayer model is also shown in figure 7(a), which shows a very similar behaviour to the SCAPS simulations, lending further credence to our developed model. There are also some other interesting observations to be made. Equations 13 and 17 predict a reduction in the bulk doping density (plateau region of 'U' shape) of these profiles with increasing thickness of the perovskite layer, due to the inverse square dependence of $N_{\text{d,min}}$ on the layer thickness $d$, as seen in figure 7(a). Additionally, we note that the apparent profiling distances in figure 7(a) and the experimental data of figure 2(c) extend to distances larger than the thickness of the perovskite layer. This is a further confirmation of the contribution of capacitances that are not related to the depletion layer capacitance, causing either an error in the values of the capacitance or the choice of relative permittivity. We will now proceed to further interpret the experimental data based on the previous analysis after a short discussion on measurement methods of these doping profiles.

Recently, Ni et al.[47] have used drive-level capacitance profiling (DLCP) to spatially resolve apparent trap densities in the perovskite solar cell. DLCP is a similar technique to a standard CV measurement but also uses a first-order term in voltage in addition to the zero-order term that gives the standard CV response.[48] The general charge response $Q$ in a capacitance measurement is given by the Taylor expansion

$$Q(\bar{V} + \tilde{V}) = \frac{d\bar{Q}}{d\bar{V}}\tilde{V} + \left(\frac{d^2\bar{Q}}{2d\bar{V}^2}\right)\tilde{V}^2 + \cdots, \qquad (18)$$

where overbars represent steady-state quantities and the tilde represents a time-dependent quantity. Equation 18 can be rearranged to give the capacitance as

$$C = \frac{d\tilde{Q}}{d\tilde{V}} = C_0 + C_1\tilde{V} + \ldots\ldots \quad (19)$$

The first term on the right-hand side of equation 19 corresponds to the zero-order capacitance term measured from CV, while the second term is included in case of a DLCP measurement. Therefore, any differences between the DLCP and CV response arise from the frequency or voltage-dependence of the first-order term's coefficient. This term has been shown to have a weak frequency dependence in perovskite solar cells, which was then used to differentiate between the free and trapped charge response going from high to low frequencies.[47] Using this assumption, Ni et al. identified a 'U'-shaped profile for both the trap and doping densities for different perovskite solar cells. We have shown previously that such an interpretation is incorrect and that equation 13 is still valid for all measurement frequencies.[29] Therefore, experimental doping profiles obtained from DLCP measurements will hereafter be shown in conjunction with CV simulations to obtain a unified and consistent interpretation.

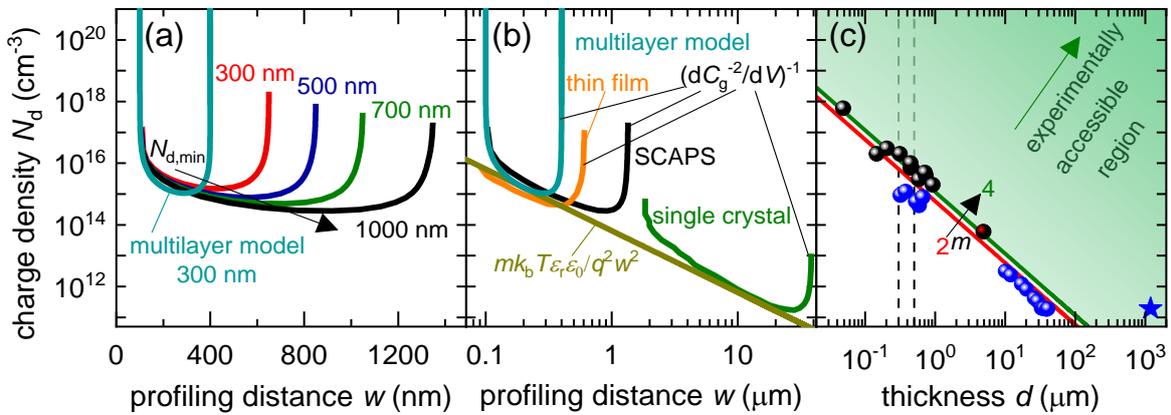

**Figure 7** (a) Simulated spatial doping profiles (from CV simulations) of a thin film perovskite solar cell for different thicknesses of the perovskite layer, using SCAPS and the multilayer model developed in section 3.1. The arrow indicates the reduction of the minimum doping density obtained in the plateau region, versus perovskite layer thickness. (b) Comparison of experimental (from ref. [47]) and simulated spatial doping profiles for different perovskite solar cells. The analytical approximation of the forward bias behaviour of the doping profile as predicted by equation 17 ($m = 1.5$ is assumed) is also shown. The slopes of the experimental data match nicely with this limit. (c) Comparison of minimum doping density resolvable from capacitance measurements and calculated minimum doping densities (lowest value of plateau region of doping profile) versus perovskite layer thickness from different capacitance measurements of perovskite solar cells. The blue data points correspond to DLCP measurements in reference [47]. The black data points correspond to CV measurements in references[22-26,42,49,50], where the perovskite layer thickness is calculated using equation S2 in the supporting information (whose validity is assumed during a Mott-Schottky analysis) with $\varepsilon_r = 30$ from the low bias saturation capacitance value. The black dotted lines indicate the typical thin-film perovskite layer thicknesses used. The area between the red and green lines sets the range of the limiting doping density value depending on the $m$ value of the capacitance transition. PSCs generally show large $m$ values, see figure 2(a). The star indicates the only data point (from ref. [47]) which lies above the resolution limit and the black point filled with red indicates a device without selective contacts (ref. [50] ).

Figure 7(b) shows a comparison of experimental doping profiles obtained from DLCP measurements of both a perovskite thin film and bulk single crystal, in conjunction with a CV simulation of a dopant-free, trap-free perovskite solar cell and a simulated doping profile from the multilayer model. The typical 'U' shape of the experimental doping profile and the reduction in apparent bulk doping density with increasing thickness is clearly observed in this figure. For small profiling distances (forward bias), figure 7(b) illustrates clearly that the experimental and simulated data follow the predicted $w^{-2}$ dependence of the apparent doping densities in equation 17. Figure 7(c) summarises the minimum doping densities obtained from capacitance-voltage (black) and DLCP (blue) measurements reported in literature and compared to the minimum doping density limit $N_{d,min}$ derived (equation 13), for different thicknesses of the perovskite layer. This limit is plotted for different $m$ values, depending on the type of capacitance that dominates the capacitance step (see figure S8 in the supporting information). The DLCP data points all lie just below the $m = 2$ line for both thin films and bulk single crystals, while the data points from CV measurements lie close to the $m = 5$ line. Only one data point lies clearly in the green region well above the resolution limit, a perovskite single crystal measured by Ni. et al.[47] Apart from this data point, all the other data points correspond to complete devices with transport layers and therefore, we conclude that the transport layer capacitances dominate the capacitance step at forward bias. In addition, figure 7(c) contains one data point (black point filled with red) that corresponds to a perovskite solar cell without selective contacts (ITO/CH$_3$NH$_3$PbBr$_3$/Au), measured by Peng et al.[50] This data point also lies very close to the resolution limit and is experimental evidence of a situation where the chemical capacitance dominates the capacitance step. We note that a significant displacement along the x-axis is also observed for some of the Mott-Schottky data when compared to typical thin-film thicknesses used (shown using black dashed lines in figure 7(c)), where the perovskite layer thickness is estimated from the saturation of the capacitance at the lowest bias point. This is related to the fact that the experimentally-observed capacitance plateau at low forward biases does not directly correspond to the geometric capacitance of the perovskite layer (see figure 5(a) and (b)) and related errors in the assumed permittivity.

In summary, the application of the traditional Mott-Schottky method to obtain doping densities and the built-in potential in the perovskite solar cell is complicated by the concomitant contributions of the transport layer geometric capacitances and the chemical capacitance of the perovskite layer to the total measured capacitance at forward bias. Therefore, depending on the $m$ value of the multilayer/chemical capacitance transition, only charge densities significantly higher than the $N_{d,min}$ for the given thickness of the perovskite layer can be considered to originate from any real doping or trapped charge densities (green region in figure 7(c)). Our analysis of experimental doping profiles from literature and our own samples indicate that the calculated doping/trap densities of perovskite solar cells using capacitance-voltage methods are below the limit of resolution provided by the fundamental response of a trap-free, dopant-free intrinsic perovskite layer.

*3.5. Capacitance plateaus: evolution versus frequency*

In this section, we will attempt to shed some light on the mechanisms governing the frequency-dependent evolution of the dark capacitance of the perovskite solar cell. In a system that consists of several different capacitances, we expect to observe several plateaus and corresponding transitions from one net capacitance to the other with a sweep of the measurement frequency. There are a number of candidate capacitances that can cause these plateaus, which have been discussed in the introduction of section 3. These include the chemical capacitance, depletion capacitance, trap capacitance and the geometric capacitances of the individual layers of the solar cell. In the case of perovskite solar cells, we also have a capacitance response from mobile ions in the bulk and accumulated ions at interfaces.[20] The timescales of response of these capacitances (which also depend on the resistances they are coupled with) usually overlaps at least to a certain degree, making it quite difficult to identify which capacitance is responding in a given measurement.

The capacitance evolution versus frequency of an ITO/PTAA/CH$_3$NH$_3$PbI$_{3-x-}$

$_yI_xBr_y$/PCBM/Ag perovskite solar cell for different applied voltages is shown in figure 8(a). At low forward biases, we observe a single plateau at high to intermediate frequencies whose magnitude increases with applied voltage, followed by a small step in the capacitance starting at ~$10^3$ Hz, shown in figure 8(b). At large forward bias, we additionally observe a low-frequency capacitance whose magnitude is very large, in the order of millifarads per cm². This low-frequency capacitance and its evolution with voltage and light intensity has been extensively investigated and while its exact mechanism is debated [51-54], it is clearly related to the density of mobile ions within the perovskite solar cell (see figure S10 in the supporting information), which is not the central issue of this work. Therefore, we will focus on the interpretation of the high and intermediate-frequency capacitance behaviour.

To understand if this complex capacitance-frequency-voltage behaviour can originate from the simple geometric capacitance transitions of the multilayer model, we carried out simulations of a dopant and trap-free perovskite solar cell using both SCAPS and the multilayer model, shown in figure 8(c) and figure 8(d) respectively. For the SCAPS simulations, the mobilities of both the selective contacts were set at $\mu = 10^{-6}$ cm²V⁻¹s⁻¹ to increase the resistance of the contact layers sufficiently. The SCAPS simulations show a similar behaviour to the experimentally measured data, showing two capacitance plateaus at high and intermediate frequencies whose magnitudes both increased with applied voltage. Simulations using the multilayer model also generate the same experimentally observed transitions with voltage and frequency excluding the ionic capacitance transition at large forward bias and low frequency. Therefore, we can conclude that the multilayer capacitance model, which is based on transitions between the geometric capacitances of the resistive selective contact layers and the absorber layer, is a credible model that can reproduce the general electronic carrier-related features of the capacitance-voltage-frequency behaviour of the perovskite solar cell.

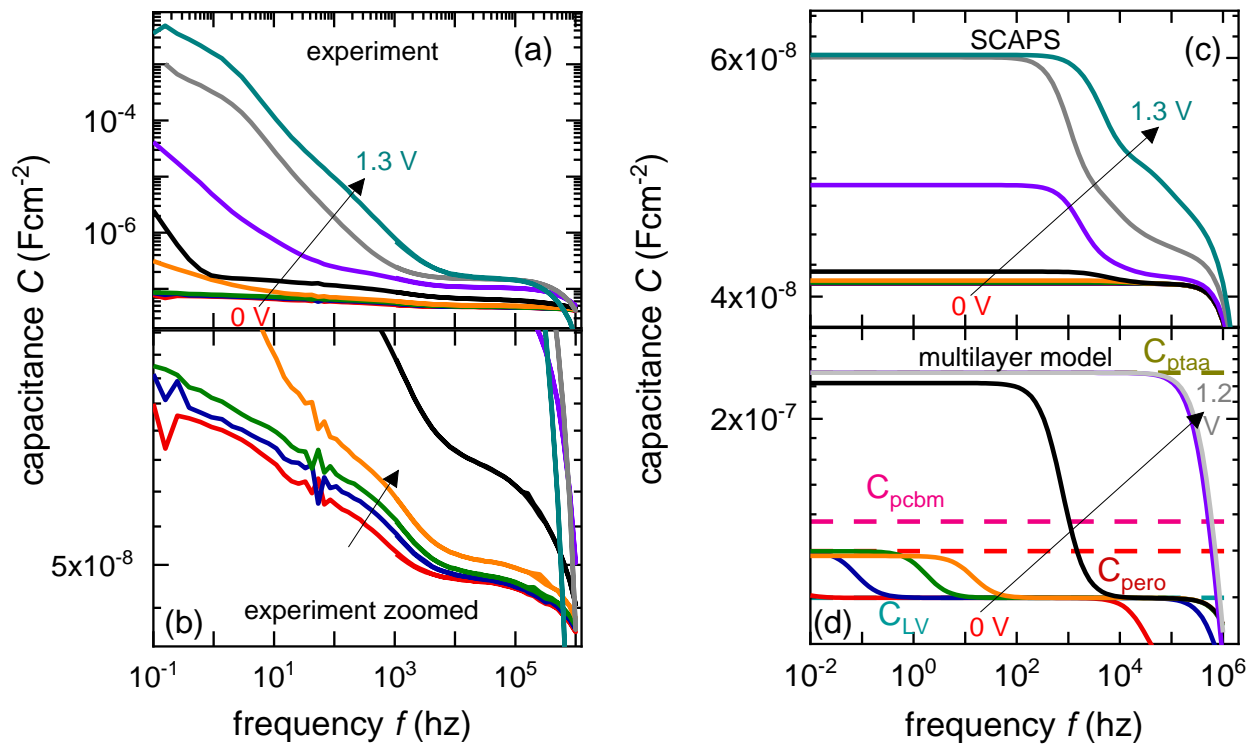

**Figure 8** (a) Experimental dark capacitance evolution versus frequency for different applied voltages for an ITO/PTAA/CH₃NH₃PbI₃₋ₓ₋yIxBry/PCBM/Ag perovskite solar cell. (b) is a zoomed plot of (a), showing an additional capacitance step (indicated by the arrow) at frequencies between $10^{-1}$-$10^4$ Hz at low forward biases. (c) SCAPS simulations of the capacitance-voltage-frequency behaviour of a perovskite solar cell (band diagram in figure 1(a)). The mobilities of the PCBM and PTAA layers were set to $\mu = 10^{-6}$ cm$^2$V$^{-1}$s$^{-1}$ in (c). (d) Simulations of the capacitance-voltage-frequency behaviour from the multilayer model. $C_{LV}$ corresponds to the low voltage capacitance plateau of equation S50 in the supporting information. The multilayer model reproduces the general capacitance features observed in experiment versus applied voltage and frequency. The steep rise in capacitance at low frequencies seen in (a) is related to the mobile ionic density in the perovskite solar cell, which are not included in the simulations of (c) and (d).

*3.6. Thermal admittance spectroscopy (TAS)*

Thermal admittance spectroscopy is a technique that aims to obtain information regarding the depth and density of trap states within the semiconductor bandgap by identifying the trap's contribution to the measured capacitance, by the variation of temperature and frequency. It has been widely used for the characterisation of different semiconductor technologies.[55-59] The method relies on the fact that the trapping and de-trapping processes occur on a characteristic timescale. This means that upon the application of a small AC voltage perturbation, the trapping-detrapping processes can either follow the modulation of the Fermi level or not depending upon the frequency employed. This leads to a frequency-dependent capacitance response from the trapped charges, which can be combined with its fundamental temperature dependence to obtain the depth of the trap within the bandgap using the relation[13] (see section A8 in the supporting information for the derivation)

$$\ln\left(\frac{\omega_{\text{inf}}}{T^2}\right) = \ln k - \frac{E_\omega}{k_B T}, \quad (20)$$

where $\omega_{\text{inf}}$ is the inflection frequency of the capacitance transition, $T$ is the temperature in Kelvin, $k$ is a constant and $E_\omega$ is the demarcation or activation energy, which is the depth of the trap energy level from the conduction or valence band. Based on equation 20, admittance measurements are made over a range of frequencies as a function of temperature, usually at zero voltage bias. From the inflection points of the capacitance versus frequency, $\omega_{\text{inf}}$ for each temperature point is determined and the slope of $(\omega_{\text{inf}}/T^2)$ versus $1/k_B T$ yields the activation energy.

Before examining the experimental data reported in literature, we will first examine the TAS response of our multilayer model. In order to determine $\omega_{\text{inf}}$ of the multilayer model, we consider two $R||C$ elements in series, yielding a transition from a high-frequency capacitance plateau to a low-frequency capacitance plateau, as shown in section A6 in the supporting information. The inflection frequency is then calculated as (see section A9 in the supporting information)

$$\omega_{\text{peak}} \propto \frac{1}{k_B T} \exp\left[-\frac{q(V_{\text{bi,TL}}-V)}{mk_B T}\right]. \quad (21)$$

Since TAS measurements are usually carried out at zero bias conditions, we then have the apparent activation energy from the multilayer capacitance model as

$$E_A = \frac{V_{\text{bi,TL}}}{m} - 3k_B T. \quad (22)$$

Simulated capacitance measurements versus frequency as a function of temperature for the multilayer capacitance model are shown in figure 9(a). We observe three plateaus, one at low temperature, followed by a transition to the second capacitance plateau at higher frequencies for increasing temperatures, while the third plateau occurs at low frequencies, below 100 Hz. The transition region for all these plateaus shifts to higher frequencies for increasing temperatures. Similar capacitance-frequency-temperature trends have been observed experimentally for the perovskite solar cell. The high-frequency transition from the first to the second capacitance plateau is referred to as the 'D1' step, while the low-frequency transition from the second to the third capacitance plateau that occurs at lower frequencies is referred to as the 'D2' step.[60] The 'D1' step is generally considered as originating from trapped charges and hence, we focus on its interpretation using the multilayer capacitance model.[60-62] Figure 9(b) shows the TAS response of the 'D1' step reported in literature for different perovskite solar cells. A large dispersion in the apparent activation energies ranging from 20 to 200 meV is calculated from these plots. The corresponding TAS response and calculated activation energies from the multilayer model for different built-in voltages of the PCBM layer are also shown (dashed lines). We note that the dominant resistance for the 'D1' transition in the multilayer model is the PCBM resistance due to the large characteristic frequency of its corresponding $R||C$ element at zero bias, see figure 5(d). Therefore, we expect the built-in voltage of the PCBM layer to affect the TAS behaviour of the 'D1' step, whose values were hence varied.

The simulated evolution from the multilayer model is quite similar to the experimental data, making a straight line at lower temperatures (large $1/k_B T$) and slightly curving downwards at higher temperatures (small $1/k_B T$) due to the temperature-dependent term in equation 22 (fits of simulated data using analytical equation 22 shown in figure S13 in the supporting information). The corresponding apparent activation energies obtained for reasonable values of the $V_{bi}$ of the PCBM layer are well within the range of commonly measured activation energies for the perovskite solar cell, as seen from the slopes of the plots in figure 9(b). This indicates that several of the experimental TAS data observed do not arise from the trapping and de-trapping of free charge carriers in the perovskite layer. Instead, they are likely an artefact caused by the contribution of the capacitances and resistances of the selective contacts. Similar conclusions were made from TAS measurements on perovskite solar cells with and without the selective contacts by Awni et al.,[60] who suggested that the 'D1' capacitance step and associated activation energy commonly measured is actually a response from trapping-detrapping processes in the hole selective contact layer.

The problem in interpretation of the TAS data becomes clear when considering equations S85 and S76 in the supporting information, that correspond to the frequency dependence of the trap capacitance and the capacitance associated with a general $R||C$ transition respectively. Both these capacitances possess an inverse square frequency dependence and hence show very similar evolution versus frequency. Therefore, the geometric capacitance transitions arising from the multilayer model must be considered as a lower limit to the measured activation energies of not only perovskite solar cells but any photovoltaic technology that makes use of additional non-metallic layers (selective contacts) in the device stack. The limit itself is determined by the electrostatic potential drop through the selective contact whose resistance dominates the capacitance transition in the frequency and temperature range of interest.

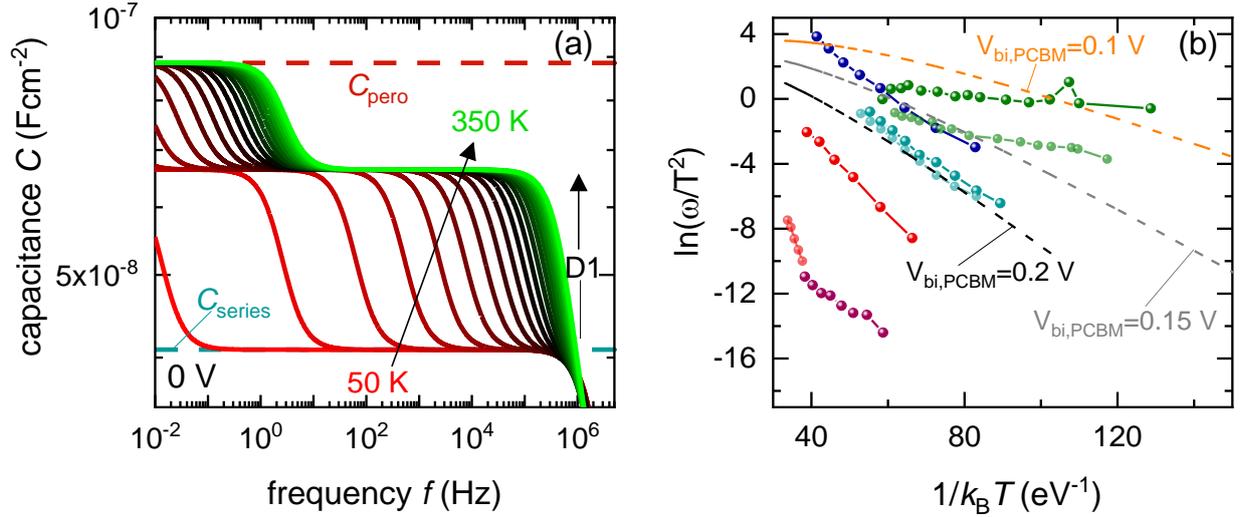

**Figure 9** (a) Simulated evolution of capacitance versus frequency for different temperatures using the multilayer model. The 'D1' step generally observed for perovskite solar cells at high frequencies is shown. (b) Comparison of thermal admittance spectroscopy (TAS) data reported in literature for perovskite solar cells with simulated TAS results using the multilayer model (dashed) for the 'D1' step using different built-in voltages $V_{bi,PCBM}$ for the PCBM layer. The literature data corresponds to references.[60-64] Fits of the simulated activation energies to the analytical solution in equation 22 are shown in figure S13 in the supporting information. Experimental data sets from the same literature reference are shown in light and dark colours. The PCBM layer mobility was reduced to $2 \times 10^{-3}$ cm$^2$V$^{-1}$s$^{-1}$ for the simulations in (a) and (b), while $V_{bi,PCBM} = V_{bi,PTAA} = 0.1$ V was used for (a).

## 4. Conclusions

We have carried out a critical assessment of various commonly-used capacitance techniques to probe perovskite solar cells. These include capacitance-voltage (CV), Mott-Schottky analysis, drive-level capacitance profiling and thermal admittance spectroscopy (TAS) measurements, which in principle yield information on key parameters such as spatial and energetic distributions of doping and trap densities. However, the use of selective contacts in these devices that are often quite resistive in nature means that these layers possess impedances whose signal can overlap with those from the perovskite layer. We explain this phenomenon by the development of a simple multilayer model that considers the perovskite solar cell as a series connection of the geometric capacitances of the selective contacts and the perovskite layer, with each capacitance in parallel with its transport or recombination resistance respectively. The voltage dependence of these resistances and hence the characteristic frequencies of each layer create a rich capacitance-voltage-frequency-temperature behaviour that is in several cases similar to that observed experimentally. This indicates that despite its simplicity, the multilayer model serves as a base model for the capacitance response of perovskite solar cells, creating fundamental limits below which the observed response cannot be considered as originating from fundamental mechanisms such as doping, trap or chemical capacitances.

In the case of Mott-Schottky measurements, such an analysis leads to a minimum doping/trap density that has an inverse dependence on the thickness of the perovskite layer. At this limit, polycrystalline, thin film solar cells will show larger apparent trap/doping densities compared to bulk, single crystal devices, which can be misconstrued as a valid result as it is intuitively expected. The multilayer model also yields a fundamental dependence of the apparent built-in voltages obtained from Mott-Schottky plots on the open-circuit voltage and measurement frequency, whose evolution is similar to that observed experimentally.[25,33,42] This means that commonly-used arguments of better charge collection due to an increased built-in voltage observed from Mott-Schottky plots upon passivation or material modification are incorrect, rather it is the higher open-circuit voltage that causes an increase in the apparent built-in voltage observed. Additionally, the model also predicts a typical 'U' shape of the spatial doping profile, yielding a constant, minimum doping/trap density in the bulk with large increases of the doping/trap densities over several orders near the perovskite/selective contact interfaces, that is also observed experimentally.[47] These large peaks in doping/trap densities are simply a consequence of the constant geometric capacitance at reverse bias and geometric capacitance transitions coupled with charge injection at forward bias. Finally, in the case of thermal admittance spectroscopy (TAS) measurements, we show that the trap-induced capacitance-frequency transition has an identical form to any arbitrary capacitance-frequency transition between $R||C$ elements in series. This yields an apparent activation energy that depends on the built-in electrostatic potential drop through the selective contact layer that dominates the transition, which must be considered as a cut-off value for the calculated trap activation energies from this method.

The calculated minimum resolution values for all these capacitance methods show a significant overlap with several experimental data sets in literature, indicating that the capacitance response of the perovskite solar cell is indeed strongly affected by its resistive contact layers.

**Author contributions**
S.R. carried out all the experiments, simulations, calculations and wrote the manuscript. Z.L. fabricated the samples. U.R. contributed to deriving the transport layer resistance (section A2 in the supporting information) and reviewing the manuscript. T.K. conceptualized, supervised the project and contributed to reviewing and editing the manuscript.

**Conflicts of interest**
There are no conflicts of interest to declare.

**Acknowledgements**
The authors acknowledge funding from the Helmholtz association via the project PEROSEED.

Electronic Supplementary Information

# Multilayer Capacitances: How Selective Contacts Affect Capacitance Measurements of Perovskite Solar Cells

Sandheep Ravishankar,[1*] Zhifa Liu[1], Uwe Rau[1] and Thomas Kirchartz[1,2]

[1]IEK-5 Photovoltaik, Forschungszentrum Jülich, 52425 Jülich, Germany

[2]Faculty of Engineering and CENIDE, University of Duisburg-Essen, Carl-Benz-Str. 199, 47057 Duisburg, Germany

## Methods

*Solar cell fabrication*

The perovskite solar cells (ITO/PTAA/CH$_3$NH$_3$PbI$_{3-x-y}$I$_x$Br$_y$/PCBM/Ag) were fabricated using the procedure detailed in ref.[1].

*Capacitance-voltage-frequency measurements*

Capacitance-voltage-frequency data were obtained from impedance spectroscopy measurements under dark conditions using a Zahner Zennium pro system. The AC perturbation used was 20 mV. The frequency range used was 10 mHz- 1 MHz.

*Simulations*

The drift-diffusion simulations without ionic densities were carried out using the program SCAPS.[2] The drift-diffusion simulations including ionic densities were carried out using the program SETFOS developed by Fluxim (www.fluxim.com). The parameters used for the simulations are provided in table S1 and S2 in the supporting information.

**Table S1** Parameters used for the SCAPS simulations, unless specified otherwise in the respective figure caption. An ITO/PTAA/CH$_3$NH$_3$PbI$_3$/PCBM/Ag perovskite solar cell was the chosen reference device upon which the simulations of the paper are based on. All simulations were made under dark conditions.

| parameter | PTAA | perovskite | PCBM |
|---|---|---|---|
| thickness (nm) | 10 nm | 300 nm | 25 nm |
| relative permittivity | 3 | 30 | 3 |
| bandgap (eV) | 3.2 | 1.6 | 2 |
| electron affinity (eV) | 2.13 | 3.93 | 4.05 |
| effective DOS CB (cm$^{-3}$) | $2 \times 10^{18}$ | $2 \times 10^{18}$ | $2 \times 10^{18}$ |
| effective DOS VB (cm$^{-3}$) | $2 \times 10^{18}$ | $2 \times 10^{18}$ | $2 \times 10^{18}$ |
| radiative recombination coefficient (cm$^3$/s) | 0 | $6 \times 10^{-11}$ | 0 |
| electron mobility (cm$^2$/Vs) | $10^{-5}$ | 20 | $10^{-1}$ |
| hole mobility (cm$^2$/Vs) | $10^{-5}$ | 20 | $10^{-1}$ |
| doping density (cm$^{-3}$) | 0 | 0 | 0 |

**series resistance**: 1 Ωcm$^2$.

**metal contacts**: The metal contact workfunctions chosen were 5.2 eV (ITO) for the PTAA layer side and 4.2 eV (Ag) for the PCBM layer side to obtain a built-in voltage of 1 V. The surface recombination velocities for electrons and holes at both metal contacts was set to $10^7$ cm/s.

**Table S2** Parameters used for the multilayer capacitance model simulations, unless specified

otherwise in the respective figure caption. All simulations were made under dark conditions.

| parameter | PTAA | perovskite | PCBM |
|---|---|---|---|
| thickness (nm) | 10 nm | 300 nm | 25 nm |
| relative permittivity | 3 | 30 | 3 |
| electron mobility ($cm^2$/Vs) | $10^{-5}$ | 20 | $10^{-1}$ |
| hole mobility ($cm^2$/Vs) | $10^{-5}$ | 20 | $10^{-1}$ |
| injection barrier at metal/layer interface (eV) | 0.288 | | 0.096 |
| built-in voltage $V_{bi,TL}$ (V) | 0.3 | 0.4 | 0.3 |
| ideality factor | | 1 | |
| electrostatic potential factor $k$ (see eq. S16) | 2 | | 2 |

**series resistance**: 1 $\Omega cm^2$.

**reverse saturation current density $j_0$**: In order to determine the reverse saturation current density using equation 4 in the main paper, a short-circuit current density $j_{sc}$ of 20 mA/$cm^2$ and an open-circuit voltage of 1.1 V under 1 sun illumination was assumed.

**Discussion of the parameters**

**Relative permittivity**: The relative permittivity value for the PTAA and PCBM layers was set to 3 since typical values for fullerenes lie between 2 and 4.[3] The value for the perovskite layer relative permittivity was chosen based on reference [4].

**Bandgap**: The PTAA layer bandgap was chosen from reference [5]. The perovskite layer considered was a $CH_3NH_3PbI_3$ perovskite and hence, its bandgap was set to 1.6 eV. The PCBM layer bandgap was chosen based on reference [6].

**Electron affinities**: The electron affinity of the PTAA layer was reduced by 100 meV from the value of reference [7] to increase the hole injection barrier at the PTAA/perovskite interface. The PCBM layer electron affinity was set to 4.05 eV based on the different values reported (3.7 and 4.2 eV[8,9]) and considering the Ag work function of 4.2 eV. The perovskite layer electron affinity was increased from 3.83 eV (obtained from reference [5]) to 3.93 eV to reduce the barrier for electrons at the perovskite/PCBM interface.

**Effective density of states (DOS)**: The effective DOS for the conduction and valence band of the perovskite layer was chosen from reference [10]. The DOS of the contact layers were chosen to be the same as that of the perovskite layer for simplicity.

**Radiative recombination coefficient**: The order of the perovskite layer radiative recombination coefficient was chosen from reference [10]. No recombination in the contact layers was assumed.

**Mobilities**: We fixed the electron and hole mobilities to be equal in all cases for simplicity. Based on the generally large mobilities reported for perovskite layers, we fixed a value of 20 $cm^2$/Vs.[11] For the PCBM layer, the value was increased by two orders from the value of reference [12] based on the fact that the Mott-Schottky plots in section 3.2 in the main paper show only two capacitance plateaus. The PTAA layer mobility was chosen based on reference [13].

**Built-in voltages and electrostatic potential factor**: The built-in voltages (based on the difference of workfunctions of the metal contacts) were divided equally between the transport layers with a slightly larger value for the perovskite layer due to its smaller capacitance. The applied external voltage was assumed to modify the electrostatic potential of the transport layers equally (i.e. electrostatic potential factor of 2 (see equation S16)), for simplicity.

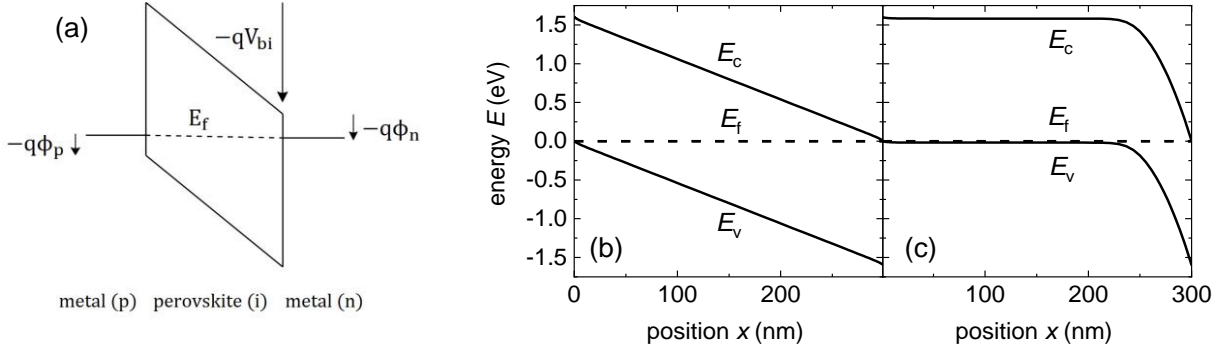

**Figure S1** (a) Schematic of an asymmetric metal (p)/perovskite (i)/metal (n) structure and band diagrams with injection barriers $\Phi_n = \Phi_p = 0$ for (b) intrinsic and (c) doped perovskite layer with acceptor density $N_A = 10^{18}$ cm$^{-3}$, used for the SCAPS simulations in figures S2 and S3. The workfunctions of the n-side metal and p-side metal were 3.93 eV and 5.53 eV respectively.

## A1. Fundamental Capacitances in Solar Cells
### A1.1. *Depletion capacitance*

The depletion capacitance is related to the existence of a region of space charge at equilibrium conditions at the interface of two semiconductors or a semiconductor and a metal. This occurs due to equilibration of the Fermi levels of the two layers upon contacting each other, which creates a region devoid of majority carriers, called the depletion region, at the interface on the semiconductor side. The total charge and hence the band-bending in this depletion region is then determined by the background doping or trap density $N_d$. Upon applying an external voltage $V$, the width of this depletion region $w$ can be modified, which means it can be considered similar to a parallel-plate capacitor with a voltage-dependent plate spacing $w$. By assuming an abrupt depletion region and uniform doping density, the Poisson equation can be solved to yield[14]

$$C_{sc} = \sqrt{\frac{q\epsilon_r\epsilon_0 N_d}{2(V_{bi}-V)}}, \tag{S1}$$

where $C_{sc}$ is the depletion or space-charge capacitance per unit area (Fcm$^{-2}$), $V_{bi}$ is the built-in voltage, $\epsilon_r$ is the relative permittivity of the semiconductor and $\epsilon_0$ is the permittivity of free space. At deep reverse bias, the semiconductor of thickness $d$ is fully depleted ($w = d$), leading to a saturation of the capacitance to the geometric capacitance value

$$C_g = \frac{\epsilon_r\epsilon_0}{d}. \tag{S2}$$

Equation S1 can be inverted to yield

$$C_{sc}^{-2} = \frac{2(V_{bi}-V)}{q\epsilon_r\epsilon_0 N_d}, \tag{S3}$$

which suggests that a plot of the inverse square of the depletion capacitance versus the applied voltage, called a Mott-Schottky plot, is a straight line whose slope contains information on the doping density and dielectric constant of the semiconductor, while the intercept on the x-axis yields the built-in voltage $V_{bi}$. Simulations of the evolution of the depletion capacitance and its corresponding Mott-Schottky plot is shown in figure S2.

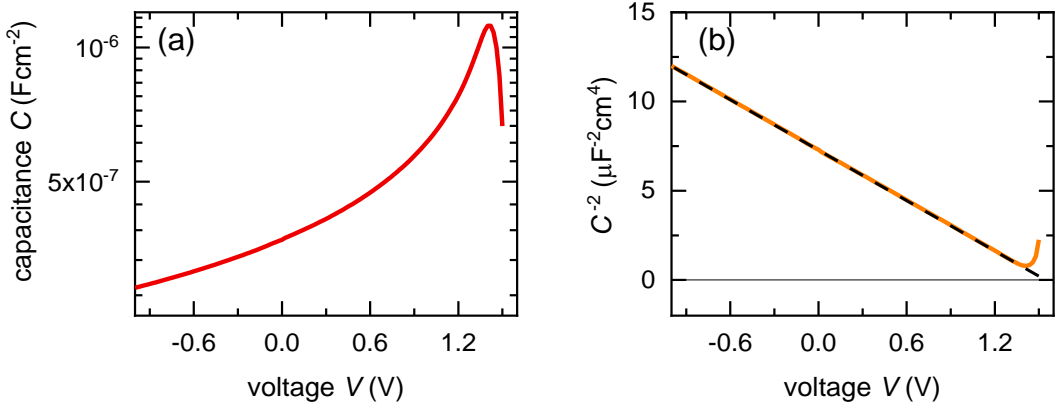

**Figure S2** SCAPS simulations of (a) evolution of the depletion layer capacitance versus voltage and (b) corresponding Mott-Schottky plot ($C^{-2}$ vs $V$). The slope of the linear Mott-Schottky region is inversely proportional to the doping density, while the intercept of the linear region on the x-axis yields the built-in potential difference as shown by the dashed line. The device considered was a metal (p)/CH$_3$NH$_3$PbI$_3$ (p)/metal (n) stack with the perovskite layer containing a uniform acceptor dopant density $N_A = 10^{18}$ cm$^{-3}$ (see figure S1(a and c)).

A1.2. *Chemical capacitance*

The chemical capacitance is related to the filling of the density of states (DOS) of the semiconductor due to a change in its chemical potential upon moving the Fermi level.[15,16] The simplest case for describing the chemical capacitance is a p-n junction with a large neutral region of thickness $d$ compared to the depletion layer width. In such a case, for a small voltage perturbation, the injected charge is simply the change in the bulk minority carrier density $n$ (since majority carriers are compensated by doping density) and we have the capacitance as[17]

$$C_\mu = q^2 d \frac{dn}{dV}. \tag{S4}$$

The bulk minority carrier concentration evolves exponentially with voltage as

$$n = n_0 \exp(\frac{qV}{mk_BT}), \tag{S5}$$

where $n_0$ is the equilibrium bulk minority carrier concentration, $m$ is a factor that controls the slope of the concentration versus voltage, $k_B$ is the Boltzmann constant and $T$ is the temperature. Combining equations S4 and S5, we get

$$C_\mu = \frac{q^2 dn}{mk_BT} = C_0 \exp(\frac{qV}{mk_BT}), \tag{S6}$$

where $C_0$ is the equilibrium chemical capacitance. For an intrinsic field-free semiconductor, equation S6 applies with $m = 2$ while for a doped semiconductor, $m = 1$. The chemical capacitance is also relevant in cases where a large built-in potential exists that separates electrons from holes, like in organic solar cells. In such cases, since the capacitance step is generated due to modification of both the electrostatic potential and the chemical potential, it may be termed as an electrochemical capacitance. This is shown in figure S3 for a metal (p)/CH$_3$NH$_3$PbI$_3$ (i)/metal (n) stack with identical injection barriers for electrons and holes (see figure S1(a),(b)).

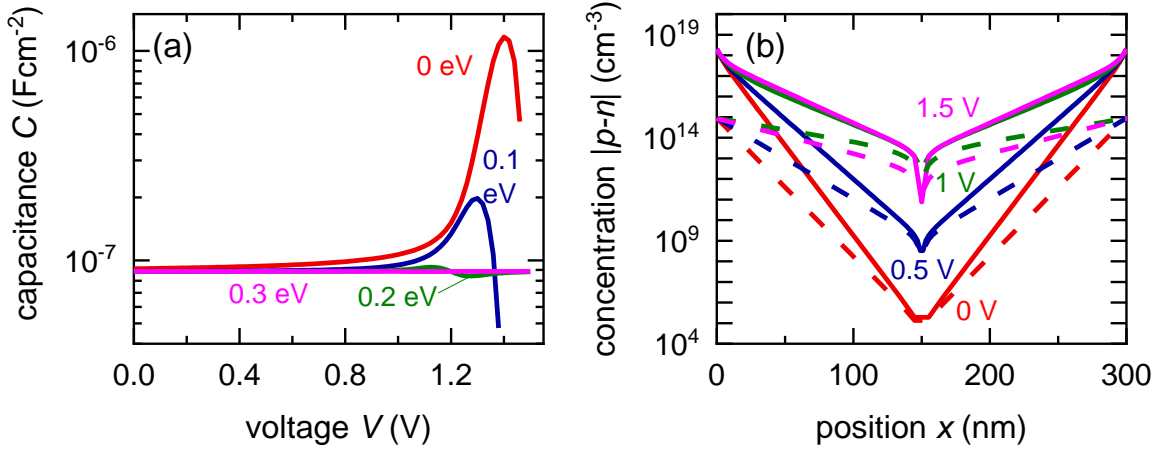

**Figure S3** (a) Simulated evolution of capacitance versus applied voltage for an asymmetric metal (p)/ $CH_3NH_3PbI_3$/metal (n) (see figure S1(a),(b)) with equal injection barriers on both sides, which are varied as shown in the captions. (b) Corresponding evolution of absolute net charge versus position for different applied voltages, for identical injection barriers of 0 eV (solid) and 0.2 eV (dashed) for electrons and holes. The exponential capacitance increase observed is a signature of the chemical capacitance, related to filling the density of states in the conduction and valence bands of the semiconductor.

Upon applying a forward bias, electrons and holes are injected into the semiconductor, causing an increase of electron and hole densities at opposite regions inside the absorber layer due to the large built-in potential difference separating the electrons and holes. This leads to an exponential increase in the capacitance. However, at large forward biases close to the built-in potential, each injected electron is balanced by a hole in the valence band through most of the absorber's thickness and hence, the net charge density and capacitance drops at this point.

Additionally, the filling of the trap density of states within the bandgap upon applying a voltage leads to a chemical capacitance of the traps. This capacitance depends on the trap occupation function that is determined by Shockley-Read-Hall (SRH) statistics.[18] The steady-state trap capacitance is given by[19]

$$C_t = \frac{q^2 N_t \Delta d}{k_B T} \overline{f}_t (1 - \overline{f}_t), \quad (S7)$$

where $N_t$ is the density of trap states, $\Delta d$ is the width of the layer in which the trap exists and $\overline{f}_t$ is the steady-state occupation function of the trap. Equation S7 suggests that the trap capacitance makes a peak at half-occupation and therefore, in principle, should be visible from a simple capacitance-voltage measurement.

A1.3. *Diffusion capacitance*

The diffusion capacitance is a closely related quantity to the chemical capacitance. It is obtained from the solution of the diffusion equation in the neutral region of a p-n junction. This capacitance is valid in the large neutral region (compared to the space-charge region) of the device, for example, seen in crystalline silicon solar cells. By solving for the AC (tilde) dark current density upon application of a small perturbation of voltage from a given steady state (overbar), we obtain[14]

$$\tilde{j} = \frac{q\tilde{V}}{k_B T} \left[ \frac{q D_p p_{n0}}{L_p \sqrt{1 + i\omega\tau_p}} + \frac{q D_n n_{p0}}{L_n \sqrt{1 + i\omega\tau_n}} \right] \exp\left(\frac{q\overline{V}}{k_B T}\right), \quad (S8)$$

where $L_p$ and $L_n$ are the diffusion lengths of holes and electrons respectively, $p_{n0}$ and $n_{p0}$ are the equilibrium minority carrier concentrations in the $n$ and $p$ layers of the p-n junction. The

capacitance can thus be obtained from the imaginary part of the admittance as

$$C_{\text{diff}} = \text{Im}[\frac{1}{\omega}\frac{q^2}{k_BT}\left(\frac{D_p p_{n0}}{L_p\sqrt{1+i\omega\tau_p}} + \frac{D_n n_{p0}}{L_n\sqrt{1+i\omega\tau_n}}\right)\exp(\frac{q\bar{V}}{k_BT})], \quad (S9)$$

whose evolution versus frequency and voltage is plotted in figure S4. Equation S9 predicts that for $\omega\tau \gg 1$, $C_{\text{diff}} \propto (\omega)^{-3/2}$. The low-frequency limit of equation S9 is given by

$$C_{\text{diff}}(\omega \to 0) = \frac{q^2}{k_BT}(\frac{L_p p_{n0} + L_n n_{p0}}{2})\exp(\frac{q\bar{V}}{k_BT}). \quad (S10)$$

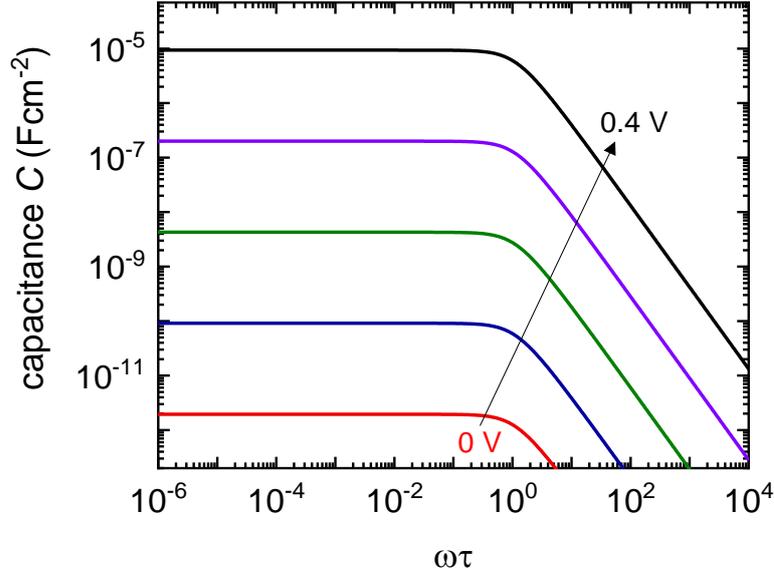

**Figure S4** Simulated evolution of diffusion capacitance versus the product of measurement frequency and lifetime for different DC forward biases using equation S9. For $\omega\tau \gg 1$, $C_{\text{diff}} \propto (\omega)^{-3/2}$. The parameters used were $n_{p0} = p_{n0} = 10^{10}$ cm$^{-3}$, $D_p = D_n = 10^{-3}$ cm$^2$s$^{-1}$, $\tau_p = \tau_n = 10^{-6}$ s.

To compare the low-frequency limit of the diffusion capacitance with that of the chemical capacitance, we consider the solution of the diffusion admittance of the base (n region) of a p-n junction of thickness $d$ with a recombination rate $S$ of the minority carriers at the surface. The low-frequency limit of the capacitance ($C_{\text{diff}-S,\text{LF}}$) calculated from the admittance solution is given by[20]

$$C_{\text{diff}-S,\text{LF}} = \left(\frac{q^2}{k_BT}\right)\frac{D_n n_{p0}}{2L_n}\left[\frac{\frac{dD_n}{L_n} - \frac{L_n dS^2}{D_n} - SL_n}{\left(\sinh^2\left(\frac{d}{L_n}\right)\right)\left(\frac{D_n}{L_n}\coth\left(\frac{d}{L_n}\right)+S\right)^2} + \tau_n\frac{\frac{D_n}{L_n}+S\coth\left(\frac{d}{L_n}\right)}{\frac{D_n}{L_n}\coth\left(\frac{d}{L_n}\right)+S}\right]\left[\exp\left(\frac{q\bar{V}}{k_BT}\right) - 1\right]. \quad (S11)$$

The limit of equation S11 when $S \to 0$, $L_n \gg d$ and at forward bias is given by

$$C_{\text{diff}}(\omega, S \to 0) = \frac{q^2 d}{k_BT}\frac{n_{p0}}{2}\exp(\frac{q\bar{V}}{k_BT}). \quad (S12)$$

Multiplying equation S12 with a factor 2 to account for the contribution of the neutral regions of both the p and n regions of the junction to the diffusion capacitance, we obtain equation S6 (assuming $n_{p0} = p_{n0}$), the chemical capacitance, with $m = 1$.

## A2. Derivation of transport layer resistance $R_{TL}$

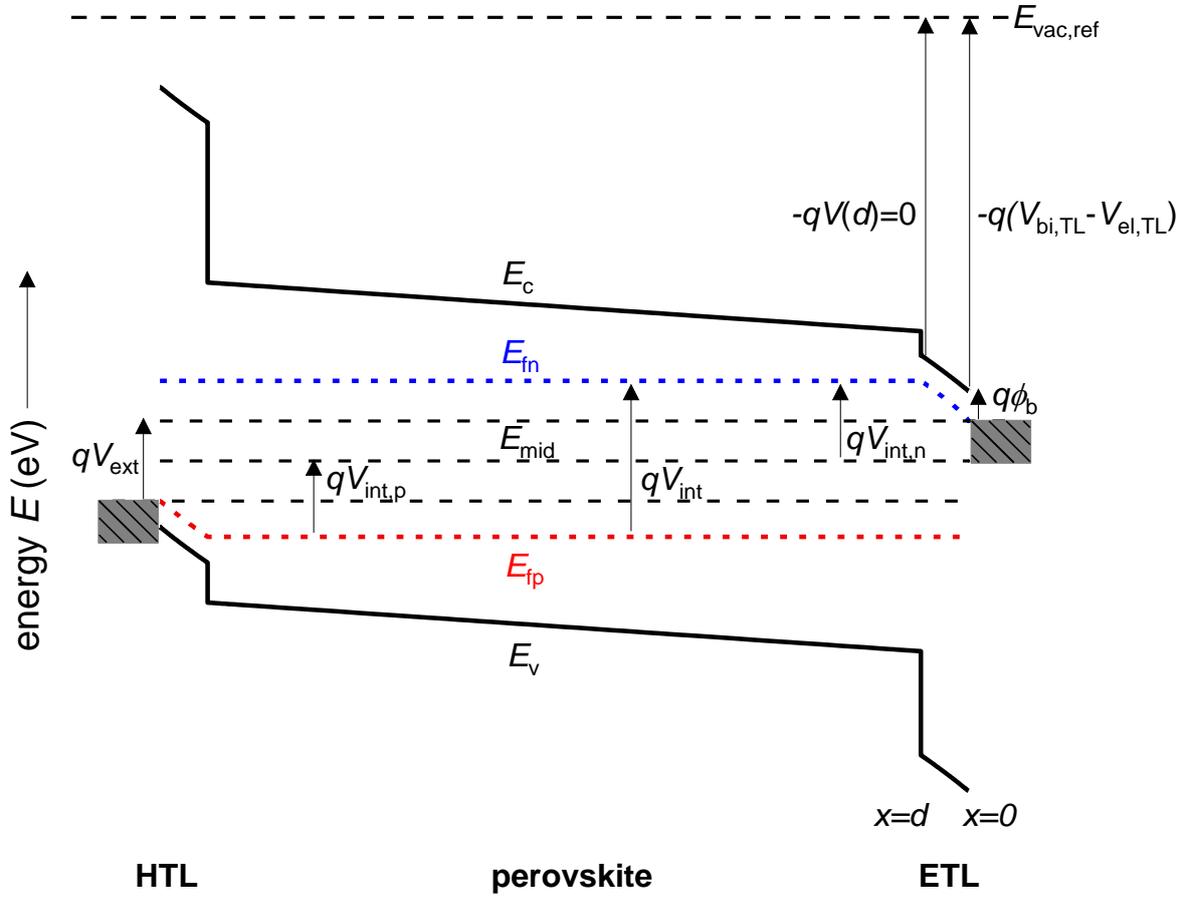

**Figure S5** Schematic of the energetics of a perovskite solar cell. The external voltage $qV_{ext}$ is applied by the metal contacts (grey boxes). $qV_{int}$ is the internal Fermi-level splitting in the perovskite layer. $\phi_b$ is the injection barrier for electrons at the electron transport layer (ETL)/perovskite interface. $E_{mid}$ corresponds to the middle of the bandgap of the perovskite. $V_{bi,TL}$ is the built-in electrostatic potential in the ETL at $x = 0$ and $V_{el,TL}$ is the amount of the applied external voltage that goes to the electrostatic potential of the electron transport layer. $E_{vac,ref}$ is the reference vacuum level with respect to which all potentials are described.

Consider the energy diagram of the perovskite solar cell in figure S5. The current through the electron transport layer (hereafter referred to as transport layer) is given by

$$j = n(x)\mu_n \frac{dE_{fn}(x)}{dx}, \tag{S13}$$

where $n$ is the electron concentration, $\mu_n$ is the electron mobility and $E_{fn}$ is the electron Fermi level. Equation S13 can be rewritten using the Boltzmann relation for the electron concentration as

$$j = N_c \exp\left(\frac{E_{fn}(x) - E_c(x)}{k_B T}\right) \mu_n \frac{dE_{fn}(x)}{dx}. \tag{S14}$$

The electric field in the transport layer is thus given by

$$F = -\frac{dV}{dx} = \frac{(V_{bi,TL} - V_{el,TL})}{d}, \tag{S15}$$

where

$$V_{\text{el,TL}} = V_{\text{ext}}/k \tag{S16}$$

and $k$ is the electrostatic potential factor, determining the amount of the external electrostatic potential that goes to the transport layer. The position of the conduction band is given by

$$\frac{dE_c}{dx} = \frac{dE_{\text{vac,s}}}{dx} = qF. \tag{S17}$$

Integrating equation S17, we get

$$E_c(x) = qFx + C, \tag{S18}$$

where $C$ is an integration constant. Substituting the value for the conduction band at $x = 0$, we have

$$E_c(x) = E_c(0) + qFx. \tag{S19}$$

Substituting equation S19 in equation S14 and rearranging, we obtain

$$\frac{j}{N_c\mu_n}\exp\left(\frac{E_c(0)+qFx}{k_BT}\right)dx = \exp\left(\frac{E_{\text{fn}}}{k_BT}\right)dE_{\text{fn}}. \tag{S20}$$

Integrating equation S20,

$$\int_0^d \frac{j}{N_c\mu_n}\exp\left(\frac{E_c(0)+qFx}{k_BT}\right)dx = \int_{E_{\text{fn}}(0)}^{E_{\text{fn}}(d)}\exp\left(\frac{E_{\text{fn}}}{k_BT}\right)dE_{\text{fn}}, \tag{S21}$$

we obtain

$$\frac{j}{N_c\mu_nF}\exp\left(\frac{E_c(0)}{k_BT}\right)\left[\exp\left(\frac{qFd}{k_BT}\right)-1\right] = \exp\left(\frac{E_{\text{fn}}(d)}{k_BT}\right)-\exp\left(\frac{E_{\text{fn}}(0)}{k_BT}\right). \tag{S22}$$

This yields the current density through the transport layer as

$$j = q\mu_n N_c F\exp\left(-\frac{E_c(0)}{k_BT}\right)\frac{\exp\left(\frac{E_{\text{fn}}(d)}{k_BT}\right)-\exp\left(\frac{E_{\text{fn}}(0)}{k_BT}\right)}{\left[\exp\left(\frac{qFd}{k_BT}\right)-1\right]}. \tag{S23}$$

From figure S5, we note the following relations

$$E_{\text{fn}}(d) - E_{\text{mid}} = qV_{\text{int,n}}, \tag{S24}$$

$$E_{fn}(0) - E_{\text{mid}} = qV_{\text{ext}}/2. \tag{S25}$$

Multiplying and dividing equation S23 by $\exp(E_{\text{mid}}/k_BT)$ and substituting for $V_{\text{int,n}}$ and $V_{\text{ext}}$ from equations S24 and S25, we get

$$j = q\mu_n N_c F\exp\left(-\frac{\frac{qV_{\text{ext}}}{2}+q\phi_b}{k_BT}\right)\frac{\exp\left(\frac{qV_{\text{int,n}}}{k_BT}\right)-\exp\left(\frac{qV_{\text{ext}}}{2k_BT}\right)}{\left[\exp\left(\frac{qFd}{k_BT}\right)-1\right]}. \tag{S26}$$

where $\phi_b$ is the injection barrier at the transport layer/metal interface. Multiplying and dividing equation S26 by $\exp(qV_{\text{ext}}/2k_BT)$, we get

$$j = \frac{q\mu_n N_c F\exp\left(-\frac{q\phi_b}{k_BT}\right)}{\left[\exp\left(\frac{qFd}{k_BT}\right)-1\right]}\exp\left(\frac{q[V_{\text{int,n}}-\frac{V_{\text{ext}}}{2}]}{k_BT}\right), \tag{S27}$$

which can be written as

$$j = j_{0,\text{TL}}\left(\exp\left(\frac{qV_{\text{TL}}}{k_BT}\right)-1\right), \tag{S28}$$

where

$$j_{0,\text{TL}} = \frac{q\mu n_0 F}{\exp\left(\frac{qFd}{k_BT}\right)-1} \tag{S29}$$

and
$$n_0 = N_c \exp(-\frac{q\phi_b}{k_B T}) \quad \text{(S30)}$$

is the electron concentration at the transport layer/metal interface. The potential drop in the transport layer is given by

$$V_{TL} = V_{int,n} - \frac{V_{ext}}{2}. \quad \text{(S31)}$$

For an intrinsic semiconductor such as the perovskite, we have $V_{int,n} = V_{int}/2$, which yields

$$V_{TL,intrinsic} = \frac{V_{int} - V_{ext}}{2}. \quad \text{(S32)}$$

Combining equations S32 and S28, we obtain the transport layer dark current as

$$j = j_{0,TL}\left(\exp\left(\frac{q[V_{int} - V_{ext}]}{2k_B T}\right) - 1\right). \quad \text{(S33)}$$

Equation S33 shows that the transport layer current depends on both the internal and external voltages and therefore, its resistance cannot simply be calculated by differentiating the current density versus the external voltage. However, the resistance of the transport layer can be obtained by solving for $V_{int}$ by equating the transport layer current in equation S33 with the dark recombination current through the device

$$j_{0,TL}\left(\exp\left(\frac{q[V_{int} - V_{ext}]}{2k_B T}\right) - 1\right) = j_0 \exp(\frac{qV_{int}}{k_B T}), \quad \text{(S34)}$$

where $j_0$ is the reverse saturation current. Equation S34 is an implicit equation that can be solved numerically using a root-finding algorithm. In order to obtain an analytical form of the transport layer resistance, we consider the situation where $V_{int} \cong V_{ext}$. This condition allows linearising the exponential in equation S33 and we therefore have

$$R_{TL} = \frac{d}{q\mu n_0 \left[\frac{q(V_{bi,TL} - V_{el,TL})}{k_B T}\right]} \left[\exp\left(\frac{q(V_{bi,TL} - V_{el,TL})}{k_B T}\right) - 1\right]. \quad \text{(S35)}$$

For the case $V_{ext} \gg V_{int}, V_{el,TL} \gg V_{bi,TL}$, we obtain

$$j = \frac{q\mu n_0}{d}(V_{bi,TL} - V_{el,TL}). \quad \text{(S36)}$$

This yields the minimum transport layer resistance as

$$R_{TL,min} = \frac{dk}{2q\mu n_0}\left(1 + k\frac{V_{bi,TL} - V_{int}}{V_{int} - V_{ext}}\right) \cong \frac{dk}{2q\mu n_0}. \quad \text{(S37)}$$

We thus have an approximate analytical equation for the transport layer resistance from the sum of equations S35 and S37 as

$$R_{TL,analytical} = \frac{d}{q\mu n_0 \left[\frac{q(V_{bi,TL} - V_{el,TL})}{k_B T}\right]} \left[\exp\left(\frac{q(V_{bi,TL} - V_{el,TL})}{k_B T}\right) - 1\right] + \frac{dk}{2q\mu n_0}. \quad \text{(S38)}$$

Figure S6 shows simulations of the calculated internal voltage and dark current versus external voltage (figures S6(a) and (b) respectively) and the steady-state, differential and analytical forms of the transport layer resistance (figure S6(c)).

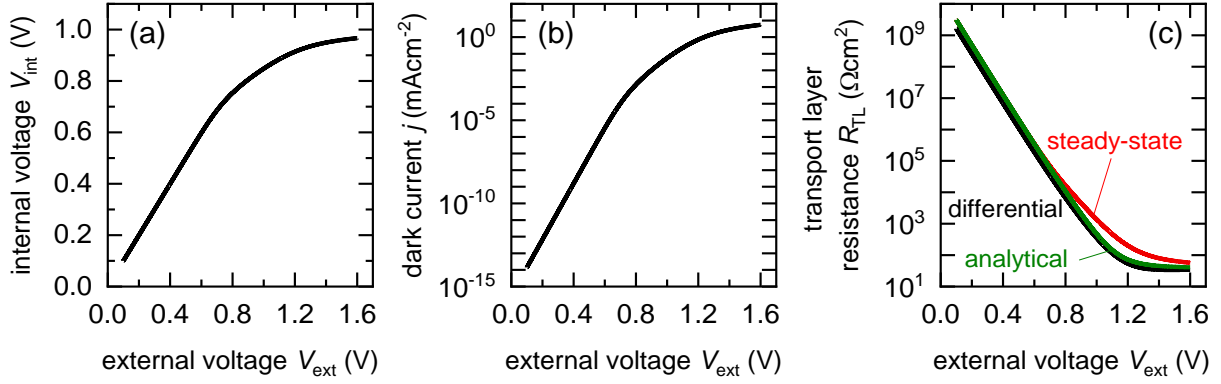

**Figure S6** (a) Simulated evolution of internal versus external voltage by numerically solving equation S34. (b) Simulated dark current density from the solution of the internal voltage and (c) Simulated steady-state, differential and analytical approximation (equation S38) of the transport layer resistance as a function of the external voltage.

## A3. Analytical approximations of capacitance plateaus

We consider the frequency-dependent capacitance transitions of figure 1(c) in the main paper based on the equivalent circuit in figure 1(b). For simplicity, we will rename these capacitances as $C_{HF}$, $C_{IF}$ and $C_{LF}$, with their corresponding resistances $R_{HF}$, $R_{IF}$ and $R_{LF}$. We assume that $C_{HF} \ll C_{IF} \ll C_{LF}$ and $R_s \ll R_{HF} \ll R_{IF} \ll R_{LF}$. The impedance of this system is given by

$$Z = R_s + (\frac{1}{R_{HF}} + i\omega C_{HF})^{-1} + +(\frac{1}{R_{HF}} + i\omega C_{IF})^{-1} + (\frac{1}{R_{HF}} + i\omega C_{LF})^{-1}. \tag{S39}$$

Equation S39 can be rewritten in terms of the respective characteristic frequencies $\omega = (RC)^{-1}$ as

$$Z = R_s + \frac{R_{HF}}{1+\frac{i\omega}{\omega_{HF}}} + \frac{R_{IF}}{1+\frac{i\omega}{\omega_{IF}}} + \frac{R_{LF}}{1+\frac{i\omega}{\omega_{LF}}} . \tag{S40}$$

*High-frequency plateau and drop-off*

To estimate the high frequency capacitance plateau, we use the condition $\omega_{HF}/\omega, \omega_{IF}/\omega, \omega_{LF}/\omega \to 0$ in equation S40 (as seen in figure 1(c) in the main paper) and we obtain

$$Z = R_s + \frac{1}{i\omega C_{HF}} + \frac{1}{i\omega C_{IF}} + \frac{1}{i\omega C_{LF}} . \tag{S41}$$

This yields the capacitance as

$$C(\omega) = \frac{C_{ser}}{1+(\frac{\omega}{\omega_{ser}})^2} , \tag{S42}$$

where $C_{ser}$ is the series combination of the three capacitances given by

$$C_{ser} = (\frac{1}{C_{HF}} + \frac{1}{C_{IF}} + \frac{1}{C_{LF}})^{-1} \tag{S43}$$

and $\omega_{ser} = (R_s C_{ser})^{-1}$. The low-frequency limit of equation S42 is given by equation S43 while the high-frequency region where $\omega \gg \omega_{ser}$ leads to a drop in the capacitance. A more rigorous derivation involves considering the condition $\omega/\omega_{HF}, \omega_{IF}/\omega, \omega_{LF}/\omega \to 0$, leading to the capacitance

$$C = [\frac{1}{C_{HF}}(\frac{R_s+R_{HF}}{R_{HF}})^2 + \frac{1}{C_{IF}} + \frac{1}{C_{LF}}]^{-1}. \tag{S44}$$

In the limit of $R_s \ll R_{HF}$, equation S44 becomes equal to equation S43.

*Intermediate-frequency plateau*

To estimate the intermediate-frequency plateau, we use the condition $\omega/\omega_{HF}, \omega/\omega_{IF}, \omega_{LF}/\omega \to 0$ in equation S40 to obtain

$$Z = R_s + R_{HF} + R_{IF} - i\omega(R_{HF}^2 C_{HF} + R_{IF}^2 C_{IF}) + \frac{1}{i\omega C_{LF}}, \tag{S45}$$

yielding the capacitance as

$$C = \left(\frac{1}{C_{LF}} + \frac{(R_s + R_{HF} + R_{IF})^2}{R_{HF}^2 C_{HF} + R_{IF}^2 C_{IF}}\right)^{-1}. \tag{S46}$$

*Low-frequency plateau*

The low-frequency plateau is obtained by taking the low-frequency limit of equation S40, which reduces the impedance to the form

$$Z = R_s + R_{HF} + R_{IF} + R_{LF} - i\omega(R_{HF}^2 C_{HF} + R_{IF}^2 C_{IF} + R_{LF}^2 C_{LF}). \tag{S47}$$

This yields the capacitance as

$$C = \left[\frac{(R_s + R_{HF} + R_{IF} + R_{LF})^2}{R_{HF}^2 C_{HF} + R_{IF}^2 C_{IF} + R_{LF}^2 C_{LF}}\right]^{-1}. \tag{S48}$$

*Low-voltage plateau*

The estimation of the low-voltage plateau in figure 5(a) in the main text requires considering the magnitudes of the resistances and characteristic frequencies of each individual layer at low forward bias, shown in figure 5(d) in the main text. We obtain the conditions $R_{pero} \gg R_{pcbm}$ and $\omega_{ptaa}/\omega, \omega_{pero}/\omega \to 0$, yielding the impedance from equation S40 as

$$Z = R_s + \frac{1}{i\omega C_{g,ptaa}} + \frac{1}{i\omega C_{g,pero}}. \tag{S49}$$

This yields the capacitance as

$$C = \left(\frac{1}{C_{g,ptaa}} + \frac{1}{C_{g,pero}}\right)^{-1}. \tag{S50}$$

*High-voltage plateau*

The estimation of the high-voltage plateau in figure 5(a) in the main text requires the conditions $R_{ptaa} \gg R_{pcbm}, R_{pero}$ and $\omega_{ptaa}/\omega \to 0$ at large forward bias, obtained from figure 5(d) in the main text. This yields the impedance from equation S40 as

$$Z = R_s + \frac{1}{i\omega C_{g,ptaa}}. \tag{S51}$$

This yields the capacitance as

$$C = C_{g,ptaa}. \tag{S52}$$

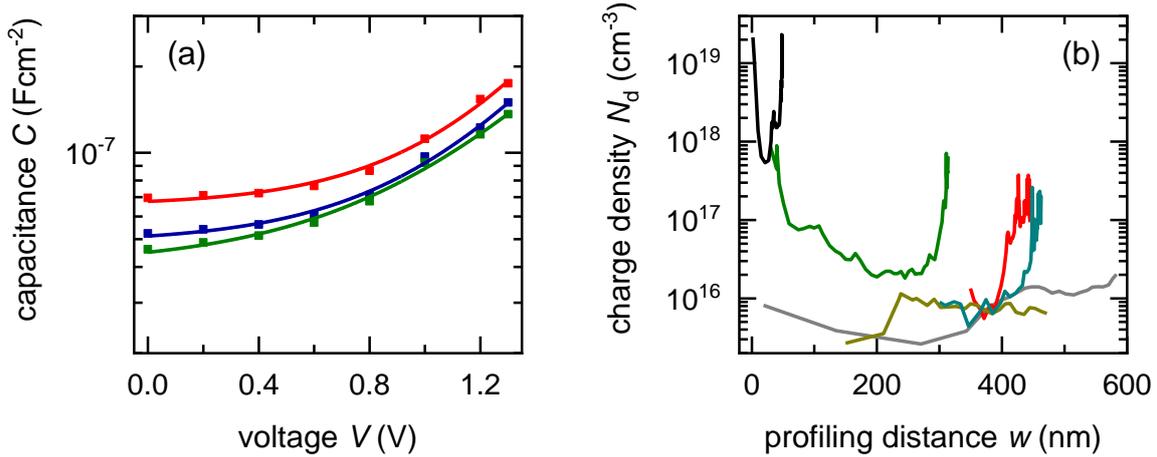

**Figure S7** (a) Fits (solid lines) of the experimental data (squares) shown in figure 2(a) in the main paper to the equation $C = C_g + C_0 \exp(qV/m k_B T)$. The fitted $m$ values are shown in the inset of figure 2(a). (b) Calculated doping profile plots of the literature data in figure 3(a) in the main text (same colours as in figure 3(a)). $\varepsilon_r = 11$ was assumed for the silicon solar cell. The doping density was estimated as the plateau with the lowest magnitude of charge density.

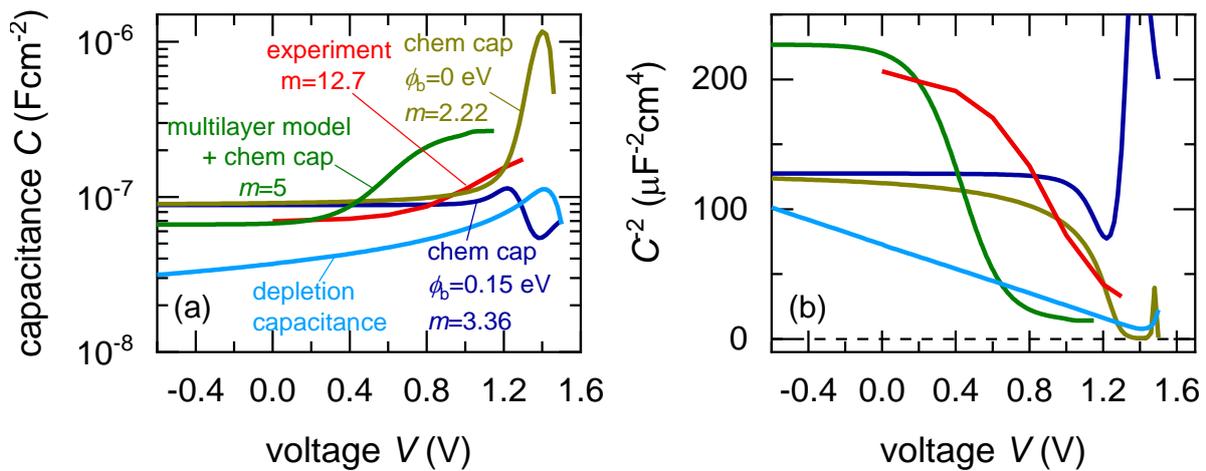

**Figure S8** (a) SCAPS simulations of the capacitance-voltage behaviour of the chemical capacitance (chem cap) for different injection barriers $\phi_b$ as shown in the labels, depletion capacitance ($N_a = 10^{18}$ cm$^{-3}$). Also shown are analytical simulations from the multilayer model including the chemical capacitance, compared to an experimental dataset. The different values of $m$ obtained from fits to the equation $C = C_g + C_0 \exp(qV/m k_B T)$ are also shown. (b) Corresponding Mott-Schottky plots of the data in (a). Discriminating the depletion capacitance from the other capacitances at forward bias in a capacitance-voltage measurement or Mott-Schottky plot is a very difficult task due to their similar evolution. The depletion capacitance was divided by a factor of 10 in (a) and its corresponding Mott-Schottky data in (b) was multiplied by a factor of 10, for ease of observation.

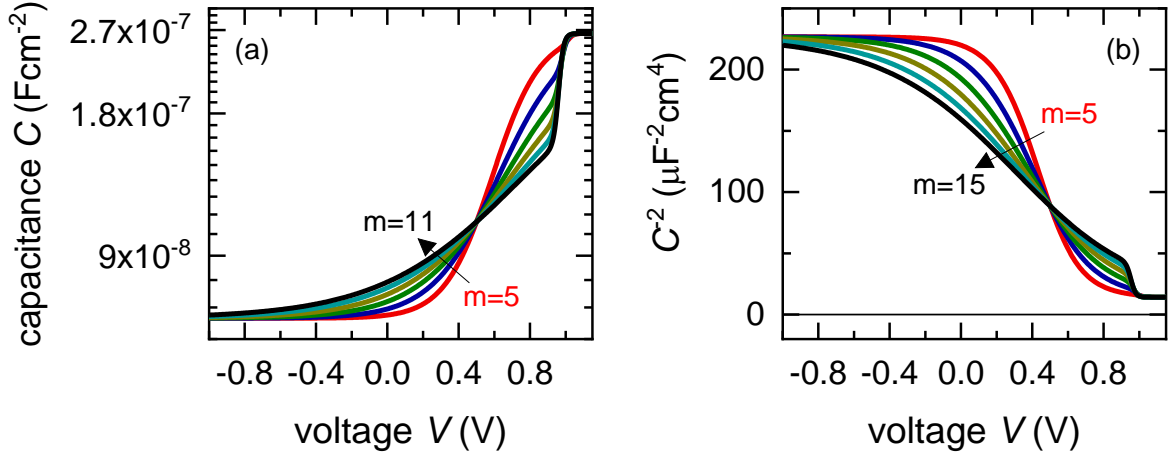

**Figure S9** Simulated evolution of (a) capacitance versus voltage and (b) Mott-Schottky plots for the multilayer capacitance model including the effect of the chemical capacitance for the perovskite layer ie: $C_{pero} = C_{g,pero}[1 + \exp(q(V - V_{onset})/k_B T)]$. $V_{onset}$ is the voltage at which the exponential rise from the chemical capacitance appears, which was set at 0.5 V based on the experimental capacitance-voltage behaviour shown in figure 2(a) in the main text. The mobility of the PCBM layer was set several orders higher than the value in table S2 to make it metal-like.

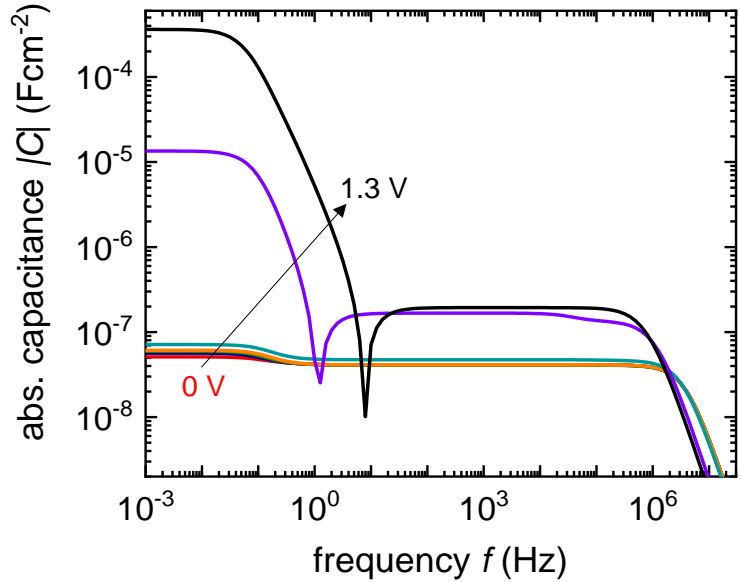

**Figure S10** Simulated absolute capacitance-voltage-frequency evolution of an ITO/PTAA/CH$_3$NH$_3$PbI$_3$/PCBM/Ag solar cell based on the parameters in table S1, including equal mobile cation and anion densities of $10^{16}$ cm$^{-3}$. The mobilities of the cations and anions was set at $10^{-9}$ cm$^2$/Vs. The large capacitance of the order of a few millifarads per cm$^2$ at low frequencies and large forward bias is a consequence of the mobile ionic densities. The kink observed at approximately 1-10 Hz is a consequence of the transition from a positive to a negative capacitance.

## A4. Analytical expression for doping profile at forward bias

This derivation was published in ref.[21] and is reproduced here for the sake of clarity. "We consider a general capacitance of the form

$$C = C_\text{g} + C_0 \exp\left(\frac{qV}{mk_\text{B}T}\right). \tag{S53}$$

The doping density profile is given by

$$N_\text{d} = \frac{-2}{q\varepsilon}\left[\frac{dC^{-2}}{dV}\right]^{-1}, \tag{S54}$$

which can be represented in terms of the slope factor $m$ of the capacitance versus voltage and profiling position $x = \varepsilon/C$ as

$$N_\text{d} = \frac{\varepsilon C}{qx^2}\left[\frac{dC}{dV}\right]^{-1}. \tag{S55}$$

For large forward bias, we have

$$C \cong C_0 \exp\left(\frac{qV}{mk_\text{B}T}\right), \tag{S56}$$

$$\frac{dC}{dV} = \frac{qC}{mk_\text{B}T}. \tag{S57}$$

Substituting equations S56 and S57 in equation S55, we get

$$N_\text{d} = \frac{mk_\text{B}T\varepsilon}{q^2 x^2}. \tag{S58}$$

Equation S58 shows that $N_\text{d} \propto x^{-2}$ at large forward bias, which explains the rise in interfacial charge densities for the lowest profiling distances. This forms the left side of the 'U'-shaped doping profile. The flat region in the profile can be described by a constant value $N_\text{d,min}$. This gives the doping profile at forward bias as

$$N_\text{d} = N_\text{d,min} + \frac{mk_\text{B}T\varepsilon}{q^2 x^2}." \tag{S59}$$

### A5. Derivation of minimum charge density for resolution

This derivation was published in ref.[21] and is reproduced here for the sake of clarity. "We again consider a general capacitance

$$C = C_\text{g} + C_0 \exp\left(\frac{qV}{mk_\text{B}T}\right), \tag{S60}$$

where $C_\text{g} = \varepsilon/d$ is the geometric capacitance of the layer of thickness $d$ and $C_0$ and $m$ are the pre-factor and slope factor respectively for the capacitance that grows exponentially with applied voltage. The doping density profile is given by

$$N_\text{d} = \frac{-2}{q\varepsilon}\left[\frac{dC^{-2}}{dV}\right]^{-1}. \tag{S61}$$

The profiling distance is given by

$$x = \frac{\varepsilon}{C}, \tag{S62}$$

and at deep reverse bias, we obtain the thickness of the layer as

$$d = \frac{\varepsilon}{C_\text{g}}. \tag{S63}$$

Substituting equations S60 and S63 in S61, we get

$$N_\text{d} = \frac{mk_\text{B}T\varepsilon C^3}{q^2 d^2 C_\text{g}^2 C_0 \exp\left(\frac{qV}{mk_\text{B}T}\right)}. \tag{S64}$$

To obtain the minimum value of the doping density, we need to solve

$$\frac{dN_\text{d}}{dx} = \frac{dN_\text{d}}{dV}\frac{dV}{dx} = 0. \tag{S65}$$

Differentiating equation S62 and S64 with respect to voltage, we obtain

$$\frac{dV}{dx} = \frac{-C^2}{\varepsilon(dC/dV)}, \quad (S66)$$

$$\frac{dN_d}{dV} = \frac{mk_BT\varepsilon}{q^2d^2C_g^2C_0}\left[\frac{3C^2(dC/dV)\exp(qV/mk_BT)-(qC^3/mk_BT)\exp(qV/mk_BT)}{\exp(2qV/mk_BT)}\right]. \quad (S67)$$

Solving equation S65 using equations S60, S66 and S67, we obtain

$$C_{\min} = 3C_0 \exp(qV_{\min}/mk_BT), \quad (S68)$$

which is the minimum value of the capacitance at a corresponding voltage $V_{\min}$. Substituting equation S68 in equation S60, we obtain

$$C_g = 2C_0 \exp(qV_{\min}/mk_BT). \quad (S69)$$

Substituting equations S68 and S69 in equation S64 at the voltage $V_{\min}$, we obtain the minimum doping density as

$$N_{d,\min} = \frac{27mk_BT\varepsilon}{4q^2d^2}. \quad (S70)$$

## A6. General $R||C$ transition

Consider two $RC$ elements in series, where each resistance is placed in parallel to its respective capacitance ($R||C$). We name the resistances $R_1, R_2$ and the capacitances $C_1$ and $C_2$ respectively. We assume that $C_1 \gg C_2$. The impedance of this system is given by

$$Z = \left(\frac{1}{R_1+i\omega C_1}\right)^{-1} + \left(\frac{1}{R_2+i\omega C_2}\right)^{-1}. \quad (S71)$$

The capacitance of this system is obtained using $C = \text{Im}(Z^{-1})/\omega$. When $R_1 \gg R_2$, the capacitance transitions from a plateau corresponding to $C_2$ at high frequencies to a plateau corresponding to $C_1$ at low frequencies, as shown in figure S11. The impedance of the transition from the high-frequency plateau to the low-frequency plateau in such conditions can be modelled as

$$Z = R_{\text{eff}} + \left(\frac{1}{R_1} + i\omega C_1\right)^{-1}, \quad (S72)$$

where $R_{\text{eff}}$ is an effective series resistance. Matching the low-frequency capacitance of equations S71 and S72, we obtain

$$R_{\text{eff}} = R_1(\alpha - 1) + \alpha R_2, \quad (S73)$$

where

$$\alpha = \sqrt{\frac{R_1^2 C_1}{R_1^2 C_1 + R_2^2 C_2}}. \quad (S74)$$

The capacitance transition can then be modelled as

$$C = C_2 + \frac{\text{Im}\left[\left(R_{\text{eff}} + \frac{R_1}{1+i\omega R_1 C_1}\right)^{-1}\right]}{\omega}. \quad (S75)$$

Solving equation S75 yields

$$C = C_2 + \frac{\left(\frac{R_1}{R_{\text{eff}}}\right)^2 C_1}{\left(\frac{R_1+R_{\text{eff}}}{R_{\text{eff}}}\right)^2 + \left(\frac{\omega}{\omega_1}\right)^2}, \quad (S76)$$

where $\omega_1 = (R_1 C_1)^{-1}$. The analytical approximation in equation S76 is also plotted in figure S11.

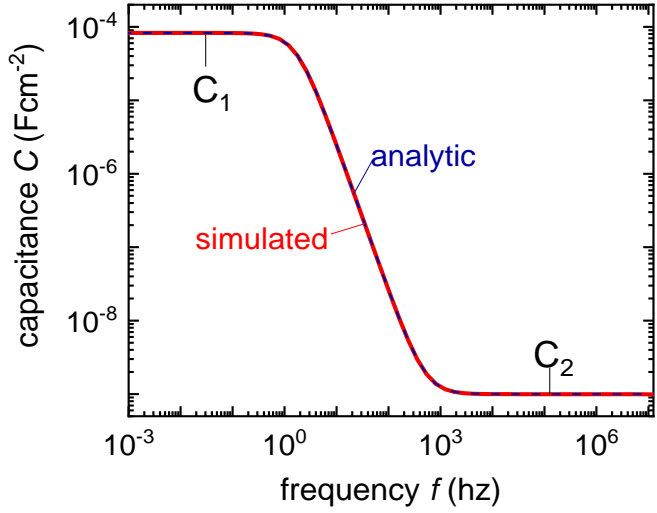

**Figure S11** Simulated evolution (red line) of the capacitance for two $R||C$ elements in series and plot of the corresponding analytical approximation equation S76 (blue dashed line).

### A7. Apparent built-in voltage of a general $R||C$ transition

Consider the general RC transition of equation S76. We set $\alpha = 1$, which gives $R_{\text{eff}} = R_2$. By assuming the perovskite recombination resistance $R_{\text{pero}}$ as $R_2$ and the transport layer resistance $R_{\text{TL}}$ as $R_1$, we obtain

$$C = C_2 + \frac{C_1}{(1+\frac{R_{\text{pero}}}{R_{\text{TL}}})^2 + (\omega R_{\text{pero}} C_1)^2}, \tag{S77}$$

where, from equations 3 and 4 in the main text, we have (assuming an ideality factor of 1)

$$R_{\text{pero}} = \frac{k_B T}{q j_{\text{sc}}} \exp\left(\frac{q(V_{\text{oc}}-V)}{k_B T}\right). \tag{S78}$$

Substituting equation S78 in equation S77, we get

$$C = C_2 + \frac{C_1}{[1+\frac{\frac{k_B T}{q j_{\text{sc}}}\exp\left(\frac{q(V_{\text{oc}}-V)}{k_B T}\right)}{R_{\text{TL}}}]^2 + (\omega \frac{k_B T}{q j_{\text{sc}}} \exp\left(\frac{q(V_{\text{oc}}-V)}{k_B T}\right) C_1)^2}. \tag{S79}$$

Equation S79 makes an apparent Mott-Schottky behaviour that shows a decreasing trend of apparent $V_{\text{bi}}$ values with decreasing $V_{\text{oc}}$ and frequency $\omega$, shown in figure S12(a). The effect of variation of frequency on the apparent $V_{\text{bi}}$ obtained from the multilayer model is also shown in figure S12(b).

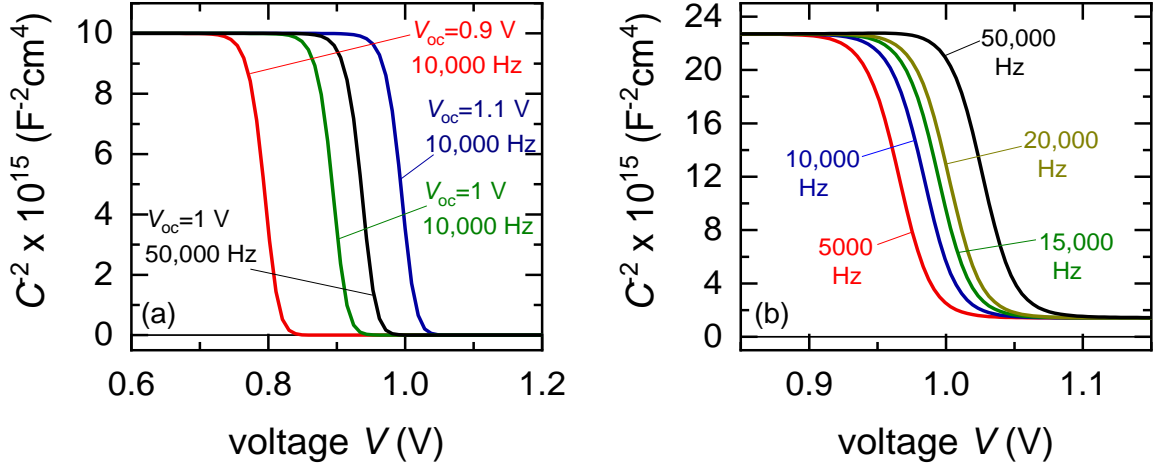

**Figure S12** (a) Simulated Mott-Schottky plots of a general $R||C$ transition (equation S79) that includes the recombination resistance of the perovskite layer for different open-circuit voltages and applied frequencies. The transport layer resistance is defined by equation 5 in the main paper. The parameters used are: $\mu_{TL} = 10^{-5}$ cm$^2$V$^{-1}$s$^{-1}$, $V_{bi,TL} = 0.3$ V, $d_{TL} = 10$ nm, $n_0 = 3 \times 10^{13}$ cm$^{-3}$, $j_{sc} = 20$ mAcm$^{-2}$, $C_2 = 10^{-8}$ Fcm$^{-2}$, $C_1 = 10^{-5}$ Fcm$^{-2}$. Higher measurement frequencies give higher apparent built-in voltages from the Mott-Schottky plot as observed experimentally in ref. [22]. Higher open-circuit voltages at the same measurement frequency yield higher apparent built-in voltages, as seen experimentally, discussed in section 3.3 in the main paper. (b) Simulated Mott-Schottky plots from the multilayer model for different applied frequencies. The PCBM layer mobility was increased by two orders of magnitude from the value in table S2. The effect of open-circuit voltage on the Mott-Schottky plot of the multilayer model is shown in figure 6(b) in the main text. The multilayer model reproduces the experimental Mott-Schottky trends observed.

## A8. Thermal admittance spectroscopy (TAS) theory

Consider the trapping and de-trapping of electrons in a density of traps $N_t$ at a single energy level $E_t$. The characteristic frequency $\omega_{td}$ of this process is given by[23]

$$\omega_{td} = \beta \bar{n} + \varepsilon = \beta \bar{n}[1 + \exp(\frac{E_t - \bar{E}_f}{k_B T})], \quad (S80)$$

where $\beta$ (cm$^{-3}$s$^{-1}$) and $\varepsilon$ (s$^{-1}$) are the rate constants of the trapping and de-trapping processes respectively, $\bar{n}$ is the steady-state electron concentration and $\bar{E}_f$ is the steady-state Fermi level. Note that equation S80 is a function of applied bias as it depends on the steady-state Fermi level and electron concentration. At equilibrium conditions, we have

$$\omega_{td,0} = \beta n_0[1 + \exp(\frac{E_t - E_{f0}}{k_B T})], \quad (S81)$$

where $n_0$ and $E_{f0}$ are the equilibrium electron concentration and Fermi level respectively. By assuming the de-trapping rate to be much slower than the trapping rate, we have the characteristic frequency as

$$\omega_{td} = \beta N_c \exp(\frac{-E_\omega}{k_B T}), \quad (S82)$$

where $N_c$ is the density of states in the conduction band and $E_\omega$ is the demarcation energy, given by

$$E_\omega = E_c - E_t. \quad (S83)$$

Note that equation S82 gives a trap frequency that is independent of the Fermi level position. Equation S82 can be inverted to yield

$$E_\omega = k_B T \ln\left(\frac{\beta N_c}{\omega}\right). \tag{S84}$$

Equation S84 shows the maximum depth of a trap from the conduction band that will respond at a given measurement frequency ω. The depth of the trap that determines the response within this demarcation energy is determined by the position of the Fermi level, since the capacitance response is dominated by the defect levels that cross the Fermi level. The capacitance response of the traps is given by[23]

$$C(\omega) = \frac{qd}{\tilde{V}} \frac{\omega_{td}}{(\omega^2 + \omega_{td,0}^2)} [\beta N_t (1 - \bar{f}) \tilde{n}], \tag{S85}$$

where $\tilde{V}$ and $\tilde{n}$ are the modulated voltage and electron concentration respectively, and $\bar{f}$ is the steady-state trap occupation probability. By scanning the measurement frequency and measuring the capacitance response, the trap density of states $N_t$ for a p-i-n type device with a built-in voltage $V_{bi}$ can be obtained as[23]

$$N_t(\omega) = \frac{-\omega V_{bi}}{q d k_B T} \frac{dC}{d\omega}, \tag{S86}$$

which can be converted to a function of the demarcation energy (depth from the conduction/valence band edge) using equation S84, if the attempt-to-escape frequency $\beta N_c$ is known. The frequency $\omega_{peak}$ of the maximum value (inflection point of the trap capacitance versus frequency) of the trap density of states is considered as the characteristic trap frequency $\omega_{td,0}$, though the inflection frequency of equation S85 is actually given by $\omega_{td,0}/\sqrt{6}$. The evolution of $\omega_{peak}$ as a function of temperature $T$ then allows the determination of the trap depth from equation S82 as

$$\ln\left(\frac{\omega_{peak}}{\sigma v_{th} N_c}\right) = \frac{-E_\omega}{k_B T}, \tag{S87}$$

where $\beta = \sigma v_{th}$, where $\sigma$ is the capture cross-section (cm²) and $v_{th}$ is the thermal velocity (cm/s) for electrons. Both $v_{th}$ and $N_c$ are functions of temperature, whose product yields a net $T^2$ temperature dependence.[24] Therefore, we have the TAS relation

$$\ln\left(\frac{\omega_{peak}}{T^2}\right) = \ln k - \frac{E_\omega}{k_B T}. \tag{S88}$$

### A9. Derivation of apparent activation energy of a general RC transition

We consider again the general RC transition of equation S76. For $\alpha = 1$, the inflection point using equation S87 is given by

$$\omega_{peak} = \frac{1}{\sqrt{3} R_2 C_1}, \tag{S89}$$

where $R_\parallel = (1/R_1 + 1/R_2)^{-1}$, the parallel combination of the two resistances. $R_\parallel = R_{TL}$, Since TAS measurements are carried out at zero applied voltage in the dark, we have $V_{int} \cong V_{ext}$ (see figure S6(a)) and $R_\parallel \cong R_{TL}$ (see figure 5(c) in the main paper). $R_{TL}$ can thus be approximated using equation S35. Substituting equation S35 in equation S89, we obtain

$$\omega_{peak} = C \times \frac{1}{k_B T} \times \left[\exp\left(\frac{q(V_{bi,TL} - V_{el,TL})}{m k_B T}\right) - 1\right]^{-1}, \tag{S90}$$

where C is a constant. Since TAS measurements are carried out zero applied voltage, we also have the exponential term being much larger than 1, which results in

$$\omega_{peak} = C \times \frac{1}{k_B T} \times \exp\left(-\frac{q V_{bi,TL}}{m k_B T}\right). \tag{S91}$$

Converting equation S91 to the TAS form of equation S88, we have

$$\ln\left(\frac{\omega_{peak}}{T^2}\right) = \ln C - \ln(k_B T^3) - \frac{q V_{bi,TL}}{m k_B T}. \tag{S92}$$

Differentiating equation S92 versus $(k_B T)^{-1}$, we obtain the apparent activation energy of the general RC transition as

$$E_A = \frac{V_{bi,TL}}{m} - 3k_BT. \tag{S93}$$

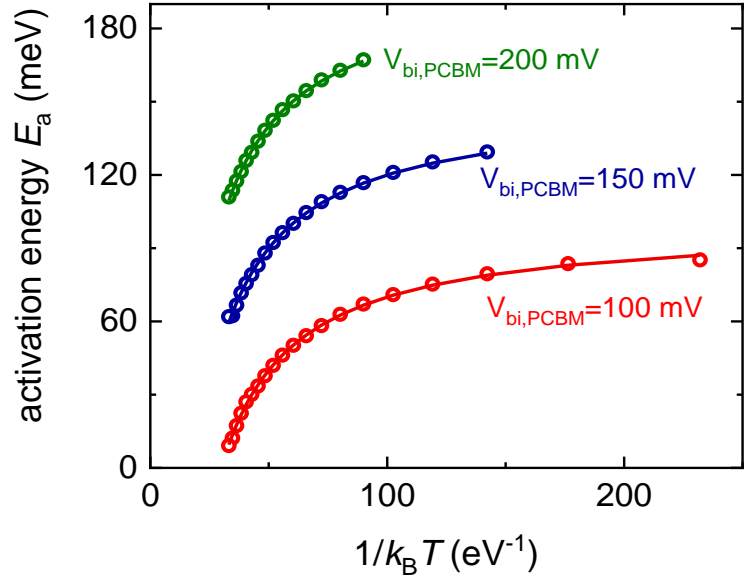

**Figure S13** Comparison of the activation energies calculated from the simulated TAS data (symbols) from the multilayer model (see figure 9 in the main paper) to the analytical form of the activation energy (lines) derived in equation S93.